\documentclass[a4paper,11pt]{article}

\usepackage{natbib}
	\setcitestyle{authoryear,open={(},close={)}, sep = {;}}
\usepackage{tikz}
\usepackage{amsmath,amssymb,amstext,amsfonts}
\usepackage{tabularx} 
\usepackage{hyperref}
	\hypersetup{
   	colorlinks=true,
   	linkcolor=red,
   	filecolor=magenta,      
	urlcolor=dgreen,
 	citecolor=blue,
	}
 \usepackage{comment}
\usepackage{subcaption}
\definecolor{dgreen}{rgb}{0.1,0.6,0.1}

\graphicspath{{FIGS/}}

\def\CTW{{\sf CTW}}
\def\BCT{{\sf BCT}}
\def\BCTs{{\sf BCT}s}

\newcommand{\pp}{\mbox{\boldmath $p$}}

\title{Change-point Detection and Segmentation of Discrete Data\\
using Bayesian Context Trees}

\author{
Valentinian Lungu 
	\thanks{Statistical Laboratory, Centre for Mathematical Sciences, 
	University of Cambridge, Wilberforce Road, Cambridge CB3 0WB, 
	UK. Email: \texttt{vml26@cam.ac.uk}.}
\and 
Ioannis Papageorgiou 	
	\thanks{Department of Engineering, University of Cambridge, 
	Trumpington Street, Cambridge CB2 1PZ, UK. Email: 
	\texttt{ip307@cam.ac.uk}.}
\and 
Ioannis Kontoyiannis 
	\thanks{Statistical Laboratory, Centre for 
	Mathematical Sciences, University of Cambridge, Wilberforce Road, 
	Cambridge CB3 0WB, UK. Email: \texttt{yiannis@maths.cam.ac.uk}.
	}
}

\begin{document}
\maketitle

\begin{center}
    \large \textbf{Abstract}
\end{center}

A new Bayesian modelling framework is introduced for 
piecewise-homogeneous variable-memory Markov chains,
along with a collection of effective algorithmic tools 
for change-point detection and segmentation
of discrete time series.
Building on the 
recently introduced 
Bayesian Context Trees (\BCT) 
framework, the distributions of different segments 
in a discrete time series are described as variable-memory Markov chains.
Inference for both the presence and location of change-points 
is then performed via 
Markov chain Monte Carlo sampling. The key observation 
that facilitates effective sampling 
is that the prior predictive likelihood 
in each segment of the data 
(averaged over all models and parameters)
can be computed exactly.
This makes it possible to 
sample directly from the posterior distribution 
of the number and location of the change-points,
leading to accurate estimates and providing
a natural quantitative measure of uncertainty
in the results.
Estimates of the actual model in each segment can 
also be obtained, at essentially no additional computational cost. 
Results on both simulated and real-world data indicate that 
the proposed methodology is robust and performs as
well or better than state-of-the-art techniques.

\bigskip


\footnotetext{Preliminary versions of some of the results
in this paper were presented at the 2022 Information
Theory Workshop~\citep{lungu-pap-K:22}.}

\newpage

\section{Introduction}

Change-point detection and segmentation
are important statistical tasks with a broad range 
of applications across the sciences and engineering. 
These tasks have been extensively studied
in the case of real-valued time series; see,
e.g., the reviews by 
\cite{aminikhanghahi:17,truong:20,williams:20}.
A significant amount of work has also been done
in addressing corresponding problems for discrete 
time series.  Among the numerous applications there, 
e.g., in biomedicine, neuroscience, health system 
management, finance, and the social 
sciences \citep{chandola:12}, the most critical 
application is probably the segmentation of genetic 
data, where the most commonly used tools are based 
on Hidden Markov Models~(HMMs).
Since their introduction for modelling heterogeneous DNA 
sequences by \cite{churchill:89,churchill:92}, they have 
also become quite popular in a wide range 
of other disciplines as well \citep{kehagias:04}.
More recently, Bayesian HMM approaches have also been 
proposed and used in practice \citep{boys:04,totterdell:17}.
In this paper we take a new Bayesian approach,
based on modelling time series with change-points
as piecewise-homogeneous variable-memory Markov~chains.

As has been often noted, the two main obstacles in the direct
approach to modelling dependence in discrete time series 
via ordinary, higher-order Markov chains,
are that: (1).~they form an inflexible and structurally poor 
model class; and (2).~the number of parameters grows 
exponentially with the memory length.
Variable-memory Markov chains provide a much richer
and more flexible model class that offers parsimonious 
and easily interpretable representations of higher-order 
chains, by allowing the memory length 
to depend on the most recently observed symbols. 
These models were first introduced 
as \textit{context-tree sources}
in the information-theoretic literature 
\citep{rissanen:83,rissanen:83b,rissanen:86},
where they have been used widely
in connection with data compression.
In particular, the development of the celebrated
Context-Tree Weighting (\CTW) algorithm 
\citep{willems-shtarkov-tjalkens:95,willems:98}
is based on context-tree sources.
Variable-memory Markov models were subsequently 
examined in the statistics literature,
initially by \cite{buhlmann:99} and \cite{buhlmann:04}, 
under the name Variable Length Markov Chains (VLMCs).

The present work is based in part on 
a Bayesian modelling framework, called Bayesian 
Context Trees (\BCTs), which was recently developed 
by \cite{BCT-JRSSB:22,ctw-isit:21,K-theory:24} 
and \cite{branch-isit:22,papag-K-BA:24} for the 
class of variable-memory Markov chains.
The \BCT\ framework 
allows for \textit{exact} Bayesian inference
(see Section~\ref{sec2}), and it has been 
found to be very effective for inference with
discrete time series. Moreover, it has also
been extended to general Bayesian mixture models for 
real-valued time series \citep{BCTAR:arxiv,BCTAR3:arxiv}.

The first contribution of this work,
in Section~\ref{sec3}, is the introduction 
of a new Bayesian modelling framework for 
piecewise-homogeneous
discrete time series.
A uniform prior is placed on the number of change-points,
and an ``order statistics'' prior is placed on their 
locations \citep{green:95,fearnhead:06}, which penalizes 
short segments to avoid overfitting. Then
each segment is described 
by a \BCT\ model,
thus defining a new class of 
\textit{piecewise-homogeneous
variable-memory chains}. 
Following \cite{BCT-JRSSB:22}, we observe that
the models and parameters in each segment can be integrated out,
and we explain how
the corresponding prior predictive likelihoods
can be computed exactly and efficiently by using 
a version of the \CTW\ algorithm.

\newpage

The second main contribution, also
in Section~\ref{sec3},
is the development 
of a new class of Bayesian methods
for inferring the number and location of change-points
in empirical data. 
A collection of
appropriate Markov chain Monte Carlo (MCMC) algorithms is introduced,
that sample directly
and efficiently from the posterior distribution 
of the number and the locations of the change-points.
The resulting approach is powerful
as it provides an MCMC approximation for the entire posterior 
distribution of interest, offering broad and insightful
information in addition to the estimates of the 
most likely change-points. Finally, in Section~\ref{sec4},
the performance of our methods~is illustrated on both simulated 
and real-world data from applications in genetics and meteorology,
where
they are found to perform at least
as well as state-of-the-art approaches.
All our algorithms are implemented in the
publicly available 
{\sf R} package \BCT\
\citep{BCT:Rv1.1},
which also contains all relevant
datasets.

\medskip

\noindent
{\bf Further connections with earlier work.}
Unlike many relevant existing methods that
only obtain point estimates and may rely in part on 
{\em ad hoc} considerations, the present methodology comes
from a principled 
Bayesian approach that provides access to the entire posterior distribution 
for the parameters of interest, in particular allowing for 
quantification of the uncertainty in the resulting estimates. The Bayesian approach was applied to the change-point detection problem in many influential works such as \cite{adams:07}, \cite{barry:93} or \cite{fearnhead:06}; 
these works consider real-valued time series and they treat the observations within each segment as independent. In contrast, our model not only allows 
Markov dependency within each segment, but also works with both long and 
short memories through the \BCT~framework.  The most closely related 
prior work is that of \cite{gwadera:08}, 
where VLMC models are used in conjunction
with the BIC criterion \citep{schwarz:78} or with
a variant of the Minimum Description Length
principle \citep{rissanen:87} to estimate variable-memory models
and perform 
segmentation by solving
an associated Bellman~equation.

There has also been a long line of works on 
change-point detection
in the information-theoretic literature.
There, piecewise-homogeneous models were first 
considered by \cite{merhav:93}
and \cite{merhavF:95},
who determined the 
optimal cumulative log-loss (or `redundancy', in the 
language of data compression).
Starting with \cite{willems:96b}, a series
of papers examined sequential change-point
detection,
typically (but not exclusively) for independent 
and piecewise identically
distributed models. These works
\citep{shamir:99,shamir:00,shamir:01,shamir:03}
are primarily concerned with deriving
theoretical bounds on the best achievable
performance by on-line methods, and also propose
sequential algorithms for change-point
detection, mostly for piecewise 
independent and identically distributed (i.i.d.)
data.  In a related but different direction,
\cite{jacob:08b} and subsequently
\cite{juvvadi:13,verma:19,yamanishi:18},
frame the change-point
detection problem as a series of hypothesis
tests, performed using statistics based 
on data compression or entropy estimation algorithms. 
Similarly, \cite{han:17}, develop optimal Bayesian change-point 
detection tests for Markov processes. Although optimal, such tests are hard to use in practice. An approach combining HMMs with information-theoretic ideas is adopted in \cite{koolen:13}, where prediction is performed on 
piecewise stationary~data.

Finally, in the context of generalisations of the original 
\CTW\ framework, change-point models have also been
examined by \cite{veness-et-al} and \cite{veness:13},
and more recently by \cite{shimada:21}, who consider
optimal and practical `Bayes codes' for data 
compression in connection with these models.

\section{Background: Bayesian context trees}
\label{sec2}

The \BCT\ framework 
of \cite{BCT-JRSSB:22}, briefly outlined in this section,
is based on variable-memory Markov chains,
a class of models that offer 
parsimonious representations of $D$th order, homogeneous 
Markov chains taking values in a 
finite alphabet 
\mbox{$A = \{0, 1, \ldots, m-1\}$}. 
The
maximum memory length $D \geq 0$ and the alphabet size $m \geq 2$ are fixed
throughout this section.
Each model describes the distribution of a process $\{X_n\}$ conditional on its initial $D$ values $X_{-D+1}^0 = x_{-D+1}^0$, where we write $X_i^j$ for a vector of random variables $(X_i , X_{i+1} , \ldots, X_j)$ and similarly $x_i^j = (x_i, x_{i+1}, \ldots , x_j)$ for a string in $A^{j-i+1}$.
This is done by specifying the conditional distribution of each 
$X_n$ given $X_{n-D}^{n-1}$. 

\begin{figure}[ht!]
    \centering
    \includegraphics[width=6cm]{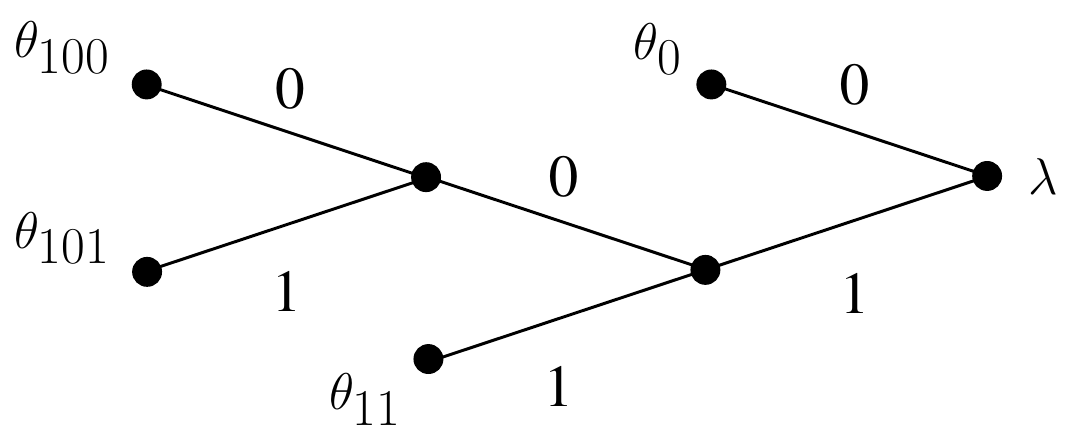}
  \caption{Tree model and parameters of a 3rd order 
	variable-memory chain.}\label{fig:tree_example}
\end{figure}

The key step in the \textit{variable-memory} model representation is the assumption that the distribution of $X_n$ in fact only depends on a (typically strictly) shorter suffix $x_{n-j}^{n-1}$ of $x_{n-D}^{n-1}$. All these suffixes, called {\em contexts}, can be collected into a proper $m$-ary tree $T$ describing the {\em model} of the chain. [A tree here is called {\em proper} if all its nodes that  are not leaves have exactly $m$ children.]
For example, consider the binary tree model on $A=\{0,1\}$ shown in 
Figure~\ref{fig:tree_example}. With $D=3$, given $X_{n-3}^{n-1}=111$, the distribution of $X_n$ only depends on the fact that the two most 
recent symbols are 1s, and it is given by the parameter~$\theta_{11}$. 
In contrast, given $X_{n-1}^{n-3} = 100$, the distribution of 
$X_n$ depends on all three recent symbols, and it is given by the 
parameter $\theta_{100}$.

Let ${\mathcal T}(D)$ denote the collection of all proper $m$-ary trees 
with depth no greater than~$D$. To each leaf $s$ of a model $T\in{\mathcal T}(D)$, 
we associate a probability vector 
$\theta_s$ that describes
the distribution of $X_n$ given that the most recently observed context is $s$:
$\theta_s=(\theta_s(0),\theta_s(1),\ldots,\theta_s(m-1))$. 

\medskip

\noindent
{\bf Prior structure.} 
The prior distribution on
models $T\in{\mathcal T}(D)$ is given by,
\begin{equation}
    \pi(T) = \pi_D(T; \beta) = \alpha^{|T|-1}\beta^{|T| - L_D(T)},
    \label{eq:prior_BCT1}
\end{equation}
where $\beta \in (0,1)$ is a hyperparameter, $\alpha = (1-\beta)^{1/(m-1)}$, 
$|T|$ is the number of leaves of $T$, and $L_D(T)$ is the number of leaves 
of $T$ at depth $D$. This prior penalizes 
larger and more complex models by an exponential amount.
Following the discussion in \cite{BCT-JRSSB:22}, 
the suggested default value of $\beta = 1 - 2^{-m+1}$ is 
used throughout this paper.

Given a tree model $T\in{\mathcal T}(D)$, an independent 
Dirichlet$(1/2, 1/2, \ldots, 1/2)$ 
prior 
is placed on the parameters $\theta = \{\theta_s: s\in T\}$ associated 
to the leaves $s$ of $T$, so that,
$
    \pi(\theta|T) = \prod_{s \in T} \pi(\theta_s) ,
$
where, 
\begin{equation}
\pi(\theta_s) = \pi(\theta_s(0), \theta_s(1), \ldots, \theta_s(m-1)) \propto \prod_{j=0}^{m-1} \theta_s(j)^{-1/2} .
 \label{eq:prior_BCT2}
\end{equation}
The extension to more general Dirichlet$(\alpha_0,\ldots,\alpha_{m-1})$
priors is straightforward \citep{BCT-JRSSB:22}
and will not be considered here.

\medskip

\noindent
{\bf Likelihood.} Given a tree model $T \in {\mathcal T}(D)$ and
associated parameters $\theta = \{\theta_s: s\in T\}$,
the induced likelihood of a time series $x=x_{-D+1}^n$ consisting
of the observations $x_1^n$ and initial context $x_{-D+1}^0$,
is given by,
\begin{equation} \label{lik_old}
   P(x|T, \theta) : =  P(x_1 ^ n | T, \theta , x_ {-D+1} ^ 0) =  \prod _{i=1} ^ n P(x_i| T, \theta, x_{-D+1}^{i-1} ) = \prod _{s \in T}\  \prod_{j=0}^{m-1} \theta_s(j)^{a_s(j)},
\end{equation}
where, for each $s\in T$,
the elements of the count vector $a_s=(a_s(0),a_s(1),\ldots,a_s(m-1))$
are:
\begin{equation}
a_s (j):=\mbox{[\# times symbol $j\in A$ follows context $s$ in $x_1^n$]},
\quad j\in A.
\label{eq:counts}
\end{equation}
Observe that~(\ref{lik_old}) is simply an instance of the Markov property.

\smallskip

\noindent \textbf{Marginal likelihood.} Given a tree 
model $T \in {\mathcal T}(D)$, it is possible to integrate out 
the parameters and compute the marginal likelihood $P(x|T)$ 
of $x=x_{-D+1}^n$ in closed form
\citep{BCT-JRSSB:22}:

\begin{equation*}
    P(x|T) :=  \int P(x_1^n|x_{-D+1}^0, T, \theta) \pi(\theta|T) = \prod_{s \in T} \frac{\prod_{j = 0}^{m-1} (1/2)(3/2)\ldots(a_s(j) - 1/2)}{(m/2)(m/2+1)\ldots(m/2+M_s-1)},
\end{equation*}
where 
$M_s = a_s(0) + a_s(1) + \cdots + a_s(m-1)$
and the counts $a_s(j)$ are 
defined in~(\ref{eq:counts}).

\medskip

\noindent {\bf Exact Bayesian inference.}
An important property of the \BCT\ framework is that it 
allows for \textit{exact} Bayesian inference. In particular, 
for a  time series 
$x = x_{-D+1}^n$,
the prior predictive likelihood (or evidence) $P_D^*(x)$ ,
averaged over both models and parameters, namely,
\begin{equation*} 
P_D^*(x) =
	\sum_{T\in {\mathcal T}(D)} 
	\pi(T) \int_{\theta} P(x|T, \theta) \pi(\theta|T) d\theta , 
\end{equation*}
can be computed \textit{exactly} by the \CTW\ 
algorithm \citep{BCT-JRSSB:22}, despite the fact
that
the number of models in ${\mathcal T}(D)$ is doubly exponential
in $D$. Moreover, the \BCT\ algorithm \citep{BCT-JRSSB:22} 
can be used to efficiently identify the
\textit{maximum a posteriori} probability~(MAP) tree model.

\section{\BCT-based change-point detection}
\label{sec3}

Here we describe the proposed Bayesian
modelling framework for piecewise-homogeneous
discrete time series,
and the associated inference methodology
for change-point detection.

Each 
segment is modelled by a homogeneous 
variable-memory chain as in Section~\ref{sec2}. 
Two different cases are considered: When the number of 
change-points $\ell$ is known $\textit{a priori}$, 
and when $\ell$ is unknown and needs to be inferred as well.
Consider a time series $x = x_{-D+1}^n$ consisting of the observations $x_1^n$ along with their initial context~$x_{-D+1}^0$. Let $\ell$ denote the number 
of change-points and let
$1 = p_0<p_1 <p_2 <\cdots<p_\ell<p_{\ell+1} = n$
denote their locations,
where we include the end-points $p_0=1$ and $p_{\ell+1}=n$
for convenience. We write,
$\pp = (p_0, p_1, \ldots , p_{\ell+1})$.

\subsection{Known number of change-points}  
\label{section:fixed}

{\bf Piecewise homogeneous Markov models.}
Suppose the maximum memory length $D$ is fixed.
Given a time series $x=x_{-D+1}^n$, 
the number of change-points $\ell$,
and their locations
$\pp = (p_0, p_1, \ldots , p_{\ell+1})$, 
the observations
$x_1^n$ are partitioned into $(\ell+1)$ segments,
\begin{align*}
   & x(1; \pp)  = x_{1}^{p_1-1},\\
   & x(j; \pp)  = x^{p_{j}-1}_{p_{j-1}}, \text{ for } j = 2, 3, \ldots, \ell ,\\
  & x(\ell+1; \pp)  = x_{p_{\ell}}^n .
\end{align*}
Each segment $x(j;\pp)$, $1 \leq j \leq \ell+1$, is 
assumed to be distributed as a 
variable-memory chain with model $T^{(j)}\in{\mathcal T}(D)$, parameter vector $\theta^{(j)}$, and initial context given by the $D$ symbols preceding $x_{p_{j-1}}$ (i.e., the last $D$ symbols of the previous segment). 
The resulting likelihood is,
\begin{equation*} 
    P(x| \pp, \{ \theta^{(j)} \}, \{ T^{(j)} \}) 
= \prod_{j=1} ^ { \ell+1}
 P( x^{p_{j}-1}_{p_{j-1}}|x_{p_{j-1}-D}^{p_{j-1}-1}, \theta^{(j)}, T^{(j)}), 
\end{equation*}
where each term in the product is 
given by~(\ref{lik_old}).

\medskip

\noindent
{\bf Prior structure.}
Given the number $\ell$ of change-points,
following \cite{green:95} and \cite{fearnhead:06},
we place a prior on their locations $\pp$ specified by 
the even-order statistics of $(2\ell+1)$ uniform draws from 
$\{2, 3, \ldots, n-1\}$ without replacement,
\begin{equation} \label{cpd_prior}
    \pi(\pp|\ell) = K_{\ell}^{-1}\prod_{j=0} ^ { \ell}  \ (p_{j+1} - p_j -1),
\end{equation}
where $K_{\ell} = \binom{n-2}{2\ell+1}$.
This prior penalizes short segments to avoid overfitting: The probability of $\pp$
is proportional to 
the product of the lengths of the segments~$x(j;\pp)$. 
Finally, given $\ell$ and $\pp$,
an independent \BCT\ prior
$ \pi(T^{(j)})\pi(\theta^{(j)}|T^{(j)})$ is placed on the tree-model and 
parameters of each segment $j$, 
as in~(\ref{eq:prior_BCT1}) and~(\ref{eq:prior_BCT2}).

\smallskip

\noindent
{\bf Posterior distribution.} 
The posterior distribution of the change-point locations $\pp$ is,
\begin{equation*}
    \pi(\pp|x) \propto P(x|\pp)\pi(\pp|\ell) ,
\end{equation*}
with $\pi(\pp|\ell)$ as in~(\ref{cpd_prior}).
 To compute the term~$P(x|\pp)$, all models~$T^{(j)}$ and parameters $\theta^{(j)}$ in
$P(x, \{ \theta^{(j)} \}, \{ T^{(j)} \} | \pp)$ need to be integrated out. 
Since independent priors are placed on different segments, 
$P(x|\pp)$ reduces
to the product of the prior predictive likelihoods,
\begin{align} 
\label{eq10}
    P(x| \pp) = \prod _ {j=1}^{\ell+1} 
  P_D^*(x(j; \pp)),
\end{align}
where the dependence on the initial context for 
each segment is suppressed  to simplify notation. 

Importantly, each term in the product~(\ref{eq10})
can be computed 
efficiently using the \CTW\ algorithm.
This means that models and parameters in each segment 
can be marginalized out, making it possible
to efficiently compute the unnormalised 
posterior distribution $\pi(\pp|x)$ 
for any $\pp$.
This is critical for effective inference
on $\pp$,
as it means that there is no need to estimate or sample 
the variables $\{T^{(j)}\}$ and~$\{\theta^{(j)}\}$ 
in order to sample directly from 
$\pi(\pp|x)$, as described~next.

\medskip

\noindent
{\bf MCMC sampler.} 
The MCMC sampler for $\pi( \boldsymbol{p}|x)$ is 
a Metropolis-Hastings \citep{robert:book} algorithm.
It takes as input
the time series $x = x_{-D+1}^n$, the alphabet size~$m\geq 2$, 
the maximum memory length $D\geq 0$, the prior hyperparameter $\beta$, 
the number~of change-points $\ell\geq 1$, the initial MCMC state 
$\boldsymbol{p}^{(0)}$, and the total number of MCMC iterations~$N$.

Given the current state $\boldsymbol{p}^{(t)} = \boldsymbol{p}$, 
a new state $\boldsymbol{p}'$ is proposed as follows: 
One of the change-points $p_i$ is chosen at random
from $\boldsymbol{p}$ (excluding the edges $p_0=1$ and $p_{\ell+1} = n$),
and it is replaced either by a uniformly chosen position $p_i'$ from 
the $(n-\ell-2)$ remaining available positions, with probability 1/2,
or by one of its two neighbours, again with probability 1/2.
Note that the neighbours of a change-point are always available,
as the prior places zero probability to adjacent change-points.

The proposed $\boldsymbol{p}'$ is either accepted and
$\boldsymbol{p}^{(t+1)} = \boldsymbol{p}'$,
or rejected and $\boldsymbol{p}^{(t+1)} = \boldsymbol{p}$, 
where the acceptance probability is
given by 
$\alpha(\boldsymbol{p}, \boldsymbol{p}') 
= \min\{1, r(\boldsymbol{p}, \boldsymbol{p}')\}$,
where,
\begin{equation*}
    r(\boldsymbol{p}, \boldsymbol{p}') 
=\frac{P(x|\boldsymbol{p}')}{P(x|\boldsymbol{p})} \times {\displaystyle \prod_{j=0}^{\ell}}\frac{(p_{j+1}' - p_j' -1)}{(p_{j+1} - p_j -1)} .
\end{equation*}
We emphasise 
the fact that the terms $P(x|\boldsymbol{p})$ 
and $P(x|\boldsymbol{p}')$ can be computed 
efficiently by the \CTW\ algorithm,
which is what enables 
this sampler to work effectively.

This sampler is designed to allow for better exploration 
of the posterior distribution
near change-points: First, the uniform jump moves of the 
sampler find a rough estimate of the location of a change-point, 
and then random-walk moves scan the region near the change-point to 
give an accurate picture of the posterior distribution around it.

\subsection{Unknown number of change-points} 
\label{sec32}

The number $\ell$ of change-points is often not known 
$\textit{a priori}$, and needs to be treated as
an additional parameter to be inferred.
As before, given the number and location of the 
change-points, each segment is modelled by a variable-memory 
chain.

\medskip

\noindent
{\bf Prior structure.} 
Given the maximum possible number of change-points $\ell_{\rm max}\geq 1$,
we place a uniform prior for $\ell$ 
on $\{0,1,\ldots, \ell_{\rm max}\}$,
and, given $\ell$, the rest of the priors 
on $\boldsymbol{p}$, $\{ T^{(j)} \}$ and $\{ \theta^{(j)} \}$ remain 
the same as in Section~\ref{section:fixed}.
The uniform prior is a common choice for $\ell$ in such
problems, as more complex models
are implicitly penalised by averaging over a 
larger number of parameters, 
resulting in what is sometimes referred to as
``automatic Occam's Razor" \citep{smith:80,kass:95,rasmussen:00}.

\medskip

\noindent
{\bf Posterior distribution.} The posterior distribution of the number 
and locations of the change-points is,
\begin{equation}
    \label{eq:variable_posterior}
    \pi(\boldsymbol{p}, \ell|x) \propto P(x|\boldsymbol{p}, \ell)\pi(\boldsymbol{p}|\ell) \pi(\ell),
\end{equation}
where $\pi(\boldsymbol{p}|\ell)$ is given in~(\ref{cpd_prior}), 
$\pi(\ell) = \frac{1}{1+\ell_{\rm max}}$, and 
the term $P(x|\boldsymbol{p}, \ell)$ is identical 
to 
$P(x|\boldsymbol{p})$ in~(\ref{eq10}).

\medskip

\noindent
{\bf MCMC sampler.}
The MCMC sampler for $\pi( \boldsymbol{p}, \ell|x)$ 
is again a Metropolis-Hastings algorithm.
Given the current state $(\ell^{(t)}, \pp^{(t)}) = (\ell, \pp)$, propose a new state $(\ell', \pp')$ as follows:
\begin{enumerate}
    \item[$(i)$] If $\ell=0$, set $\ell' = 1$, choose $p_1'$ uniformly 
	in $\{2,\ldots,n-1\}$, 
	and form $\pp'=(1,p'_1,n)$.
    \item[$(ii)$] If {$1\leq \ell < \ell_{\rm max}$}, then select
	one of the following three options with probability 1/3
	each:
    \begin{enumerate}
        \item Set $\ell' = \ell-1$ and form $\pp'$ by deleting a uniformly 
	chosen change-point from $\pp^{(t)}$;
        \item Set $\ell' = \ell+1$,
	choose a new change-point uniformly 
	from the $(n-\ell-2)$ available positions,
	and let $\boldsymbol{p}'$ be the same as $\pp^{(t)}$
	with the new point added;
        \item Set $\ell' = \ell$ and propose $\boldsymbol{p}'$ as 
	in the sampler of Section~\ref{section:fixed}.
    \end{enumerate}
    \item [$(iii)$]  If $\ell = \ell_{\rm max}$, then select one
	of the following two options with probability $1/2$ each:
    \begin{enumerate}
        \item Set $\ell' = \ell-1$ and form $\pp'$ by deleting
	 a uniformly chosen change-point from $\pp^{(t)}$;
        \item Set $\ell' = \ell$ and propose $\boldsymbol{p}'$  as 
	in the sampler of Section~\ref{section:fixed}.
    \end{enumerate}
\end{enumerate}
Finally,
either accept $(\ell', \pp')$ and set 
$(\ell^{(t+1)},\pp^{(t+1)}) = (\ell', \pp')$, or reject it and 
set $(\ell^{(t+1)},\pp^{(t+1)}) = (\ell, \pp)$, with acceptance 
probability $\alpha((\ell, \pp), (\ell', \pp')) 
= \min \{1, r((\ell, \pp), (\ell', \pp'))\}$,
where, in view of~(\ref{eq:variable_posterior}), 
the ratio $r((\ell, \pp), (\ell', \pp'))$ 
can again be computed easily via the \CTW\
algorithm; the exact form of 
$r((\ell, \pp), (\ell', \pp'))$ is
given in Appendix~\ref{app:ratio}.

\section{Experimental results} 
\label{sec4}

In this section, we evaluate
the performance of the proposed \BCT-based methods 
for segmentation and change-point detection
described in the previous section,
and we compare them
with other state-of-the-art approaches
on simulated data and  real-world time series 
from applications in genetics and meteorology. 

The only free parameters in the application
of the \BCT-based methods are the model
prior parameter $\beta\in(0,1)$, the maximum memory 
length $D\geq 0$, and the maximum number of 
change-points $\ell_{\rm max}\geq 1$.
In the experiments in this section we use the
recommended \citep{BCT-JRSSB:22} default
value of $\beta=1-2^{-m+1}$. 
The memory length~$D$ is selected by the modeller,
depending on the length of the data and the prior 
understanding about the statistical nature
of the observations;
we use the values $D=3,5$ and 10.
Finally, $\ell_{\rm max}$ should generally
be chosen at least as large as the maximum
possible number of change-points 
potentially present in the data.

In our experiments it was found
that the results were quite insensitive
to the particular choices for these three
parameters, as long as their
values were not rather obviously unreasonable.
Further discussion about the finer points of the 
choice of $\beta$ and $D$ is given in
\cite{BCT-JRSSB:22}, and about 
the choice of $\ell_{\rm max}$ in 
Section~\ref{section:maximum_number}.

\subsection{Known number of change-points}

For the simpler problem when the number of change-points
is known {\em a priori}, 
here we examine the performance of the 
\BCT-based change-point detection method of 
Section~\ref{section:fixed} on a standard
benchmark data set from statistical genetics.

The Simian virus 40 (SV40) is one of the most extensively studied 
animal viruses. Its genome is a circular double-stranded DNA molecule 
of 5243 base-pairs \citep{reddy:78}, available 
as sequence NC\_001669.1
at the GenBank database \citep{genbank:16}.
The expression of SV40 genes is regulated by two major transcripts 
(early and late), suggesting the presence of a single major change-point 
in the entire genome
\citep{churchill:92,totterdell:17,rotondo:19}.
One transcript is responsible for 
producing structural virus proteins while the other one
is responsible for producing two T-antigens (a large 
and a small one).
Hence, we take the start of the gene producing the large 
T-antigen to be the ``true" change-point of the dataset, 
corresponding to position $p_1^*=2691$.

\begin{figure}[!ht]
    \centerline{\includegraphics[width=6.8cm]{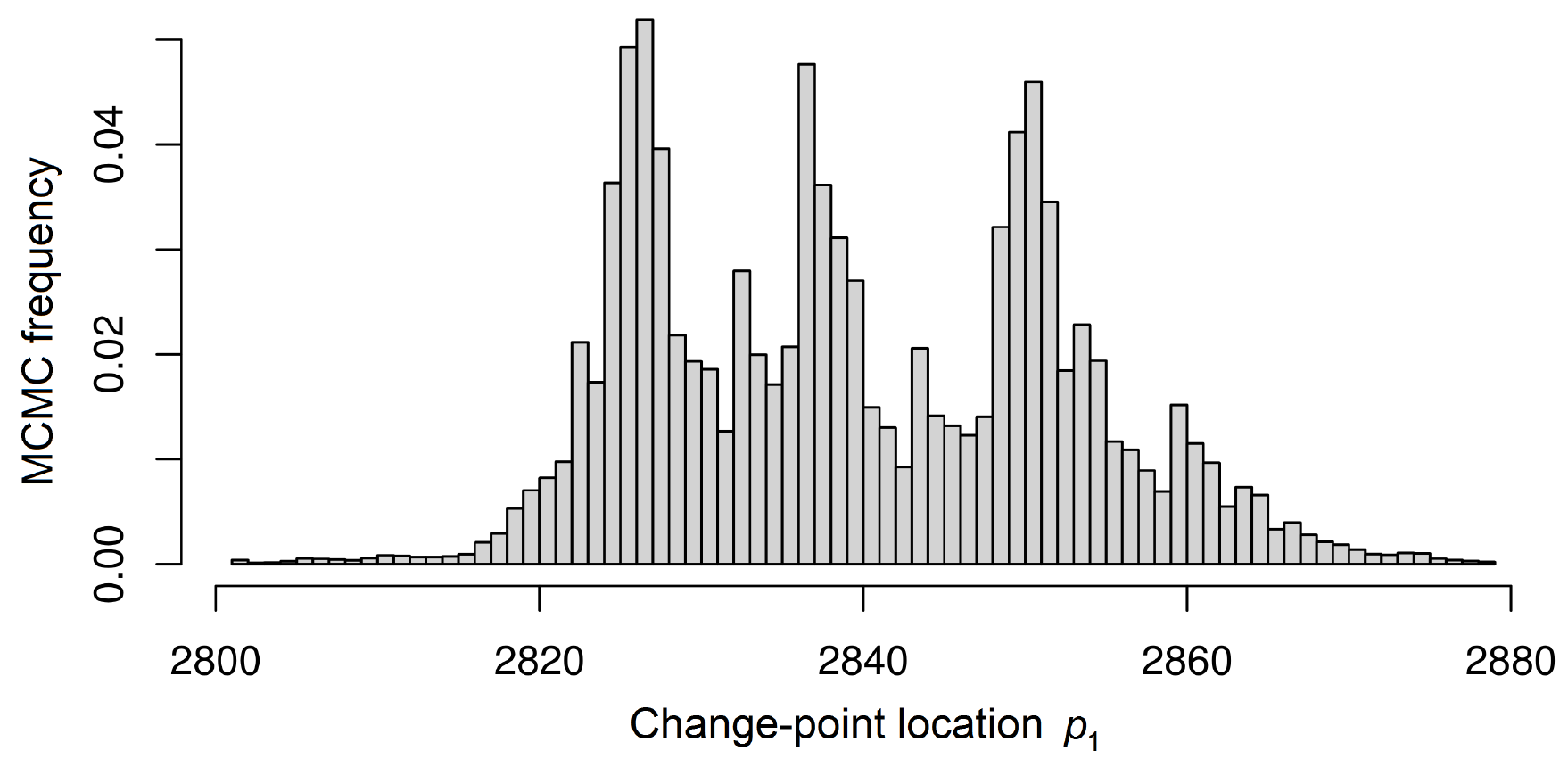}
    \includegraphics[width=6.6cm]{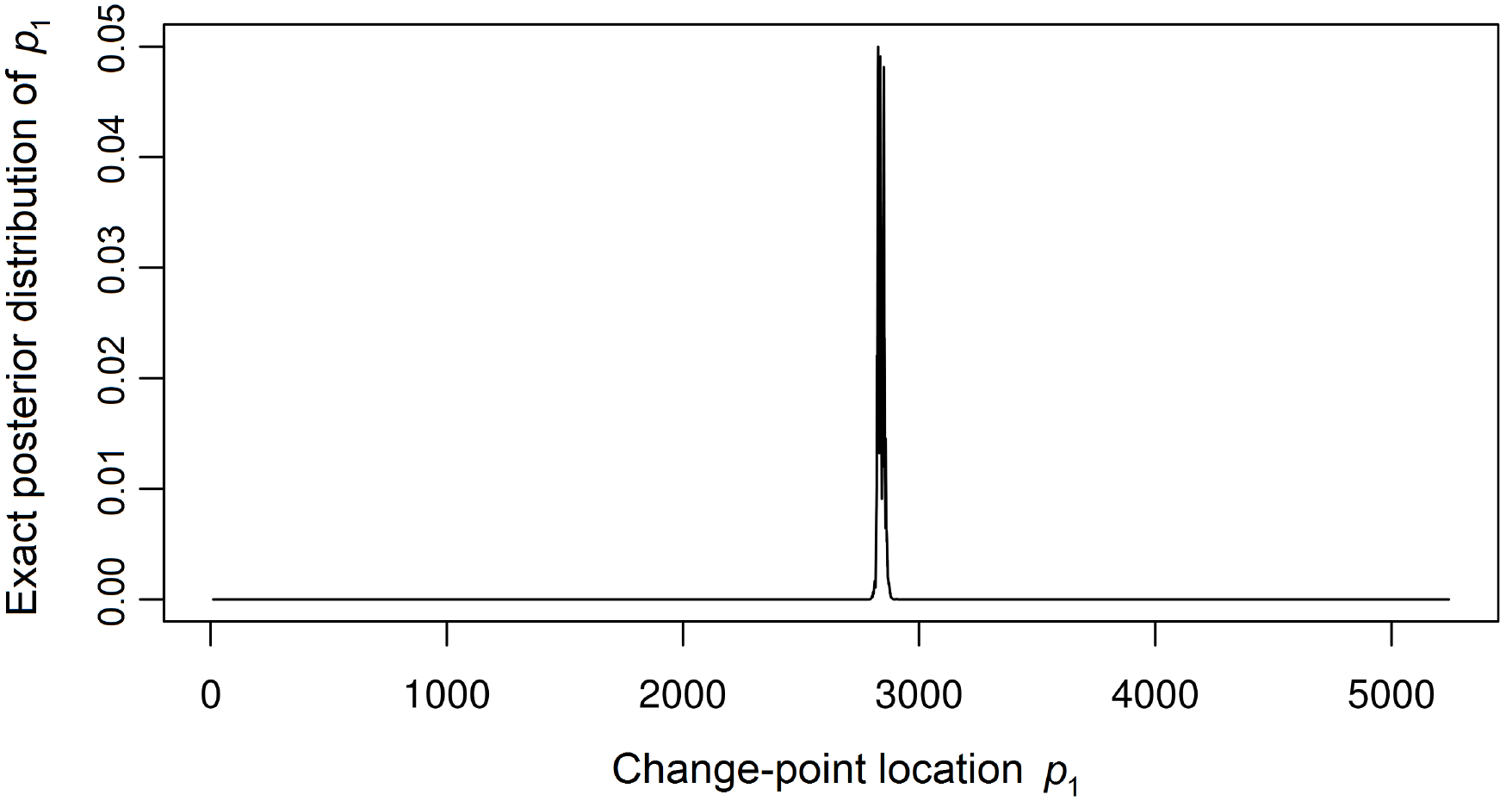}}
\caption[Posterior distribution of the change-point 
location for the SV40 genome]{Posterior distribution $\pi(p_1|x)$ of the change-point 
location for the SV40 genome. 
Left: MCMC histogram of $\pi(p_1|x)$ near its mode
based on $N =300,000$ MCMC samples, with
the first 30,000 discarded as burn-in. 
Right: Entire exact posterior distribution, $\pi(p_1|x)$.}
	\label{fig:SV1}
\end{figure}

The MCMC sampler of Section~\ref{section:fixed}
was run with $\ell=1$ and $D=10$.
The left plot in Figure~\ref{fig:SV1} shows the 
resulting histogram of the change-point $p_1$,
between locations 2800 and 2880; there were
essentially no MCMC samples outside that interval.
Viewing this as an approximation to the 
posterior distribution $\pi (p_1 | x)$, the approximate MAP location 
$\widehat p_1 =2827$ is obtained.

Compared to the results of two of the
standard, state-of-the-art methods, 
our estimate $\widehat p_1$ is slightly closer 
to $p_1^*$ than that produced
by the HMM-based technique of \cite{churchill:92} 
which gives $\widehat p_1\approx 2859$, 
and that of the Bayesian HMM approach by
\cite{totterdell:17} 
which gives~$\widehat p_1\approx 2854$.
Moreover, the \BCT-based algorithm is 
a general-purpose method that is agnostic 
with respect to the nature of the observations;
unlike many earlier approaches, it does not 
utilise any biological information about the 
structure of the data. 

\begin{figure}[!ht]
	\medskip
    \centerline{\includegraphics[width=6.8cm]{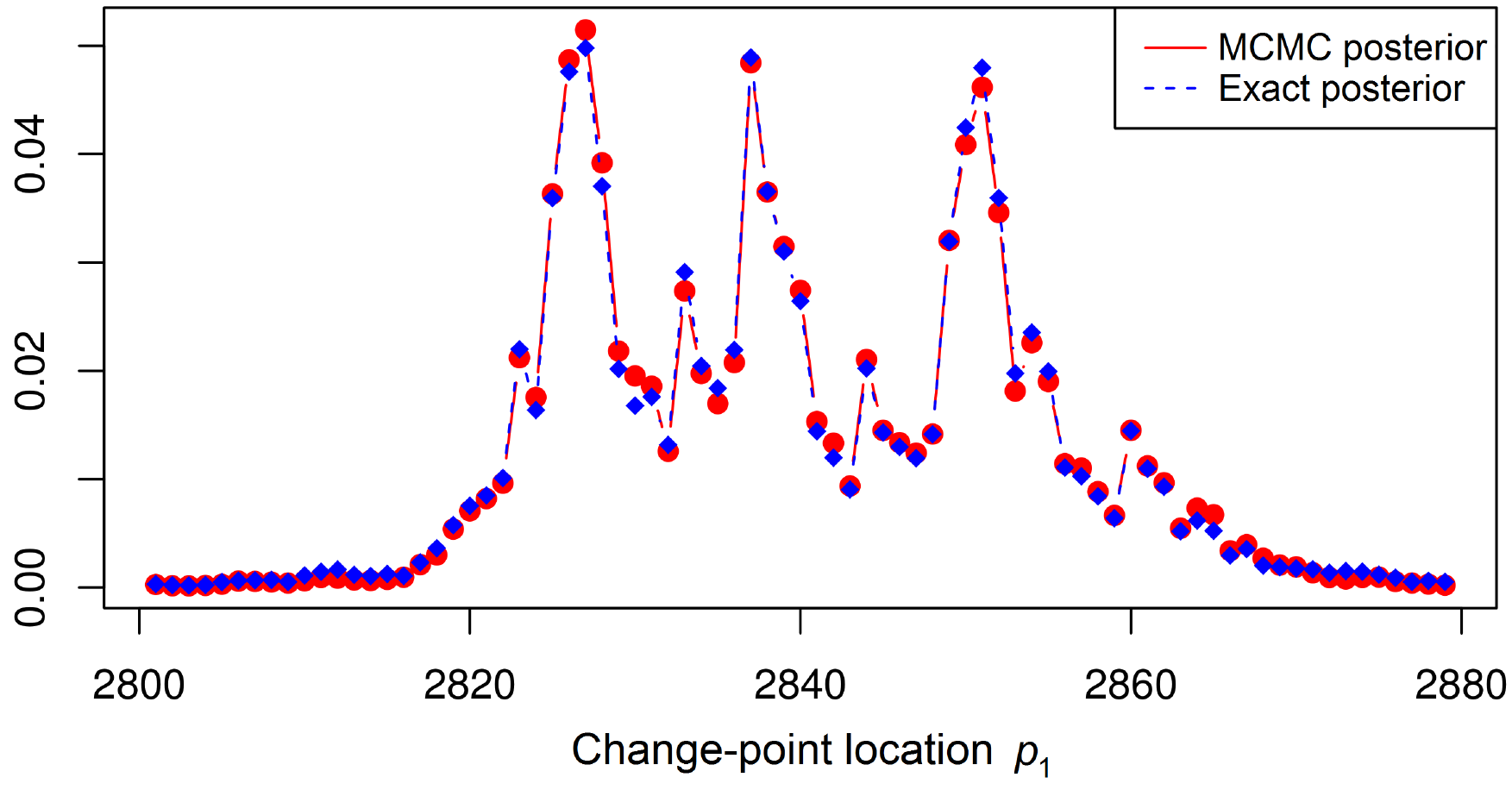}}
  \caption{Exact distribution vs.\ MCMC histogram
	of $\pi(p_1|x)$ based on $N=300,000$ MCMC samples,
	for the SV40 genome.}
	\label{fig:SV2}
\end{figure}

Since this dataset is relatively short 
and contains a single change-point, it is actually 
possible to compute the entire
posterior distribution of the location $p_1$ exactly, through,
$$\pi(p_1|x) = \frac{P(x|p_1) \pi(p_1|\ell=1)}
{\sum _{p=2} ^ {n-1} P(x|p) \pi(p|\ell=1)}.$$
Here, $\pi(p_1|\ell=1)$ is given by~(\ref{cpd_prior}), 
and the terms $P(x|p_1)$ and $P(x|p)$ 
in the numerator and denominator can be computed 
through $n-2$ applications of
the \CTW\ algorithm.
The resulting exact posterior distribution is
shown in the right plot in Figure~\ref{fig:SV1} and in more detail 
in Figure~\ref{fig:SV2}.
Even though the shape of the posterior distribution around the mode 
is quite irregular, the MCMC estimate is almost identical to the 
true distribution.
This indicates 
that the MCMC sampler converges quite fast and explores all
of the 
support of $\pi(p_1|x)$ effectively. 

In addition to a point estimate for $p_1^*$,
the posterior distribution also 
identifies an interval in which $p_1^*$ likely lies, 
illustrating a common advantage of the 
Bayesian approach. We also note that,
the fact that the posterior distribution here can be computed 
without resorting to MCMC, does not simply indicate that 
MCMC sampling is sometimes unnecessary. It actually
highlights the power of the marginalization provided 
by the \CTW\ algorithm in that, for relatively short 
datasets with few change-points, it is possible to 
obtain the entire {\em exact} posterior distribution. 

\subsection{Unknown number of change-points}

In this section we illustrate the application
of the \BCT-based change-point detection
method of Section~\ref{sec32} for the more interesting
case when the number of change-points is {\em a priori}
unknown,
and we compare its performance with existing
state-of-the-art techniques.
In Sections~\ref{section:FDR}--\ref{section:comparison} 
we examine various aspects of the change-point detection
problem using appropriately simulated data,
and in Sections~\ref{s:bl} and~\ref{s:EN} we
look at real-world datasets
from genetics and meteorology.

\subsubsection{False alarms}
\label{section:FDR}

An important aspect of change-point detection tasks is the 
potential occurrence of false positives, that is, instances where 
a change-point is erroneously detected in data where no genuine 
change exists. 
Here we investigate how the \BCT-based method behaves 
with homogeneous data. We consider three simple generating 
distributions: i.i.d.\ data
uniformly distributed over $\{0,1,2,3\}$, 
i.i.d.\ Bernoulli data with parameter $0.2$, 
and data generated by a binary variable-memory 
Markov chain with contexts ``$0$'', ``$10$'' and ``$11$''
and associated parameters:
$P(1|0)=1-P(0|0)=0.8$, $P(1|10)=1-P(0|10)=0.1$, and 
$P(1|11)=1-P(0|11)=0.5$.
The total number of observations was varied from $n=75$ to $n=1000$,
and the algorithm was ran 
with a maximum depth $D = 3$ and $\ell_{\rm max} = 2$. The posterior 
distribution of the number of change-points 
is displayed in Table~\ref{table:FDR}. 

\begin{table}[ht!]
\begin{tabular}{c|cccc}
\multicolumn{1}{l|}{} 
	& \multicolumn{4}{c}
	{Posterior distribution of the number of change-points}\\ 
\hline
\multicolumn{1}{c|}
{$n$} 
	& 75 & 100 & 500 & 1000 \\ 
\hline
i.i.d.\ uniform & ($\mathbf{0.67}$, 0.25,  0.08) 
	& ($\mathbf{0.79}$, 0.14, 0.07) 
	& ($\mathbf{0.96}$, 0.03, 0.01) & ($\mathbf{0.98}$, 0.02, 0) \\
i.i.d.\ Bernoulli & ($\mathbf{0.70}$, 0.20, 0.10) 
	& ($\mathbf{0.82}$, 0.15, 0.03) & ($\mathbf{0.90}$, 0.08, 0.02) 
	& ($\mathbf{0.95}$, 0.05, 0) \\
Markov chain
	& ($\mathbf{0.70}$, 0.20, 0.10) & ($\mathbf{0.85}$, 0.08, 0.07) 
	& ($\mathbf{0.97}$, 0.02, 0.01) & ($\mathbf{0.99}$, 0.01, 0)
\end{tabular}
\caption{The posterior distribution of the number of change-points 
for three simulated datasets. Each entry contains the vector of 
estimated posterior 
probabilities $\pi(\ell|x)$ of having $\ell=0$,~1, or~2 change-points.
In all cases, $N = 10^4$ MCMC iterations were performed, with the first 2,000 samples being discarded as burn-in.}
\label{table:FDR}
\end{table}

The results indicate that the \BCT-based algorithm
correctly assigns most of the posterior mass to $\ell=0$ 
change-points and, as the number of observations grows, 
the posterior distribution
$\pi(\ell|x)$ becomes more confident in the absence of change-points. 

\subsubsection{The choice of 
\texorpdfstring{$\ell_{\rm max}$}{}}
\label{section:maximum_number}

Here we examine how the behaviour of the \BCT-based method changes
when the
parameter $\ell_{\rm max}$ is varied. 
The first dataset we consider is made up of three segments 
of length 100 samples each; the first segment consists 
of i.i.d.\ Bernoulli data with parameter $0.8$, 
the second is generated by the simple binary 
variable-memory chain in the previous section,
and the third contains i.i.d.\ Bernoulli samples 
with parameter 0.5.
The second dataset we consider has the same distribution,
but each segment now contains 300 observations.

The results of the 
algorithm executed with $D = 3$
and $1\leq\ell_{\rm max}\leq 4$
are given in Table~\ref{table:l_max}.
They show that, if $\ell_{\rm max}$ is 
smaller than the true number of change-points $\ell ^*=2$, 
the posterior distribution places all its mass on 
$\ell = \ell_{\rm max}$,
whereas if $\ell_{\rm max}\geq\ell^*$ then
it concentrates on the true value $\ell = \ell ^ *$
and gives zero or very small probability 
on strictly smaller values of $\ell$.
Note that,
in contrast with \cite{fearnhead:06}, \cite{adams:07} or 
\cite{barry:93}, we place a uniform 
prior on the number of change-points. Nevertheless,
the posterior distribution does not overfit and is able to capture 
the correct number of change-points.

\begin{table}[ht!]
\centering
\begin{tabular}{c|cccc}
\multicolumn{1}{c|}{} & \multicolumn{4}{c}{Posterior distribution of the number of change-points} \\ \hline
\multicolumn{1}{c|}{$\ell_{\rm max}$} & 1 & 2 & 3 & 4 \\ \hline
dataset 1 & (0,$\mathbf{1}$) & (0, 0.07, $\mathbf{0.93}$) & (0, 0, $\mathbf{0.6}$, 0.4) & (0, 0, $\mathbf{0.51}$, 0.35, 0.14) \\
dataset 2 & (0,$\mathbf{1}$) & (0, 0, $\mathbf{1}$) & (0, 0, $\mathbf{0.75}$, 0.25) & (0, 0, $\mathbf{0.72}$, 0.22, 0.06)
\end{tabular}
\caption{The posterior distribution $\pi(\ell|x)$ of the number 
of change-points, estimated from the two datasets in 
Section~\ref{section:maximum_number},
based on $N = 10^4$ MCMC iterations with the first 2,000 
samples being discarded as burn-in.
Each entry in the table corresponds to $\pi(\ell|x)$ for
$\ell= 0, 1, \ldots, \ell_{\rm max}$.}
\label{table:l_max}
\end{table}
    
Next, we examine the performance of the \BCT-based method
when $\ell_{\rm max}$ is significantly larger 
than the true number of change-points.
For the first dataset, taking $\ell_{\rm max}=25,100,250,293$, 
and initialising the sampler at 
the worst possible value, $\ell^{(0)}=\ell_{\rm max}$,
the mode of the posterior distribution is still 
clearly at the
correct value $\ell=2$.
Very similar results are obtained on the 
the second dataset with $\ell_{\rm max}=10,150,300,600$; 
see Figures~\ref{fig:lmax_data2} and~\ref{fig:lmax_data1} 
in Appendix~\ref{s:ellmax}.

Table~\ref{table:running_time_lmax1} 
in Appendix~\ref{section:running_time_tables} 
shows the behaviour of running time as 
a function of $\ell_{\rm max}$. 

\subsubsection{A `difficult' simulated data set} 
\label{sec421}

We examine a synthetic dataset 
of length $n=4300$ samples, consisting 
of four different 
segments generated from four variable-memory 
chains with values in $A=\{0, 1, 2\}$;
the corresponding tree models are shown in 
Figure~\ref{fig:tree_models}.
The associated parameters, given in Appendix~\ref{app:params},
are chosen to be quite similar, so that the resulting segmentation 
problem is `difficult'. The locations of the three change-points are 
$p_1 ^ * =2500$, $p_2 ^ * = 3500$, $p_3 ^ * = 4000$. 

\begin{figure}[!ht]
\centerline{\includegraphics[width=5.4in]{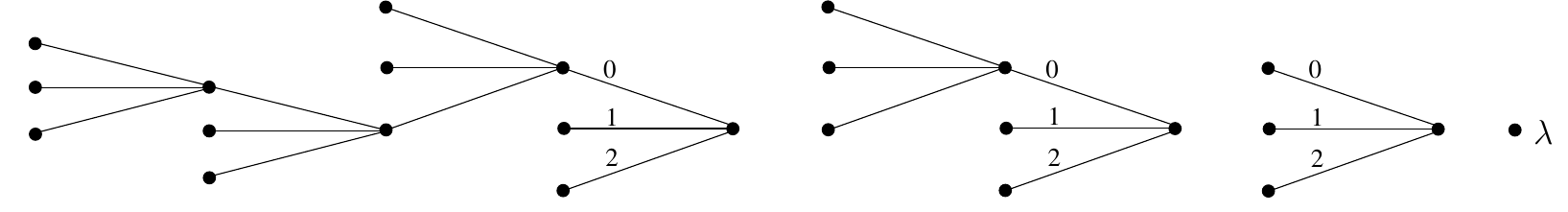}}
\caption{Four tree models of variable-memory
Markov chains on $A=\{0,1,2\}$ used for generating 
simulated data in Section~\ref{sec421}.
The last model
is the empty tree consisting of only the root
node $\lambda$, corresponding to i.i.d.\
data.}
\label{fig:tree_models}
\end{figure}

\begin{figure}[!ht]
\hspace{0.8in}
\includegraphics[height=4cm]{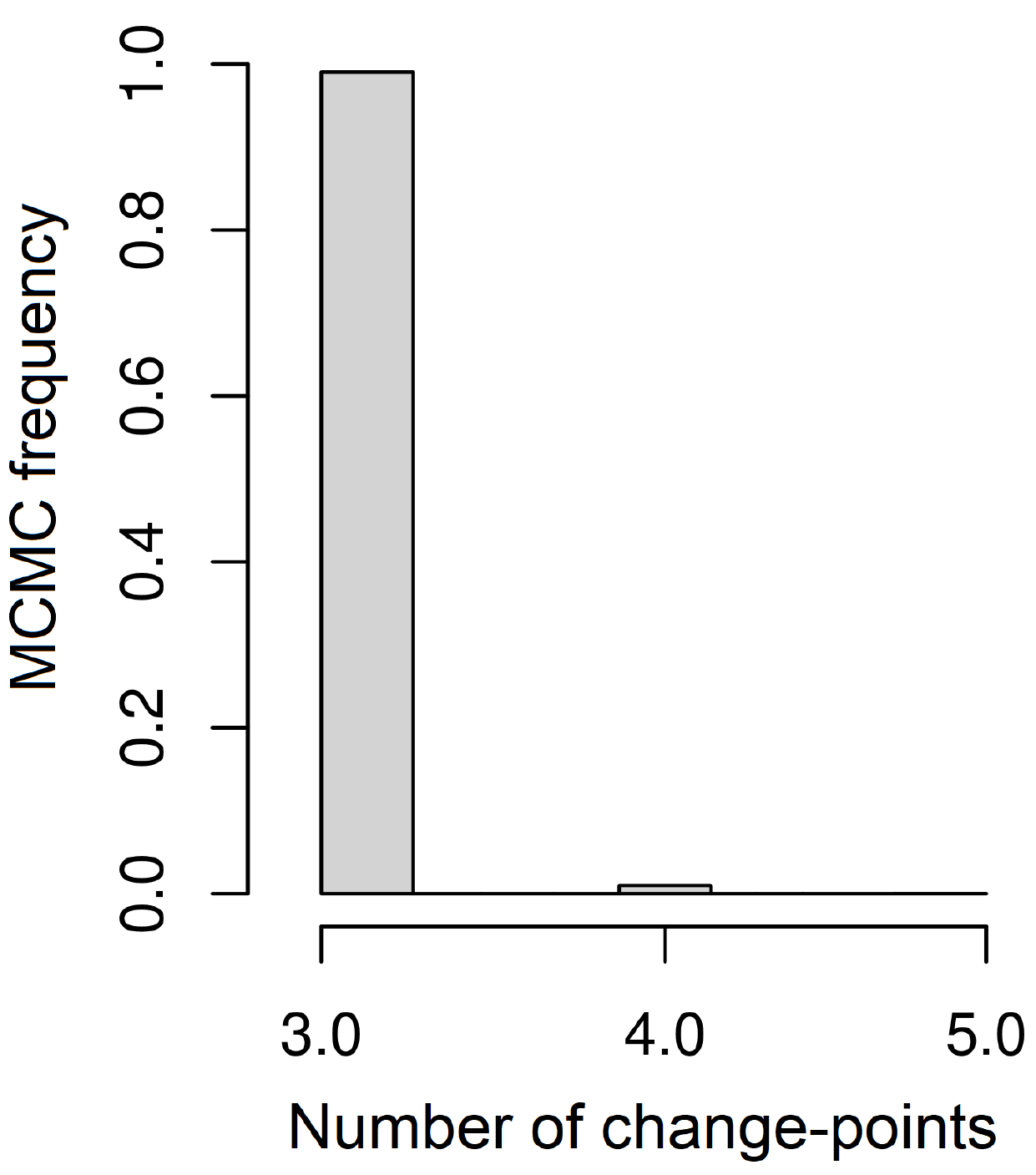}

\vspace*{-1.55in}

\hspace{2.4in}
\includegraphics[height=4cm]{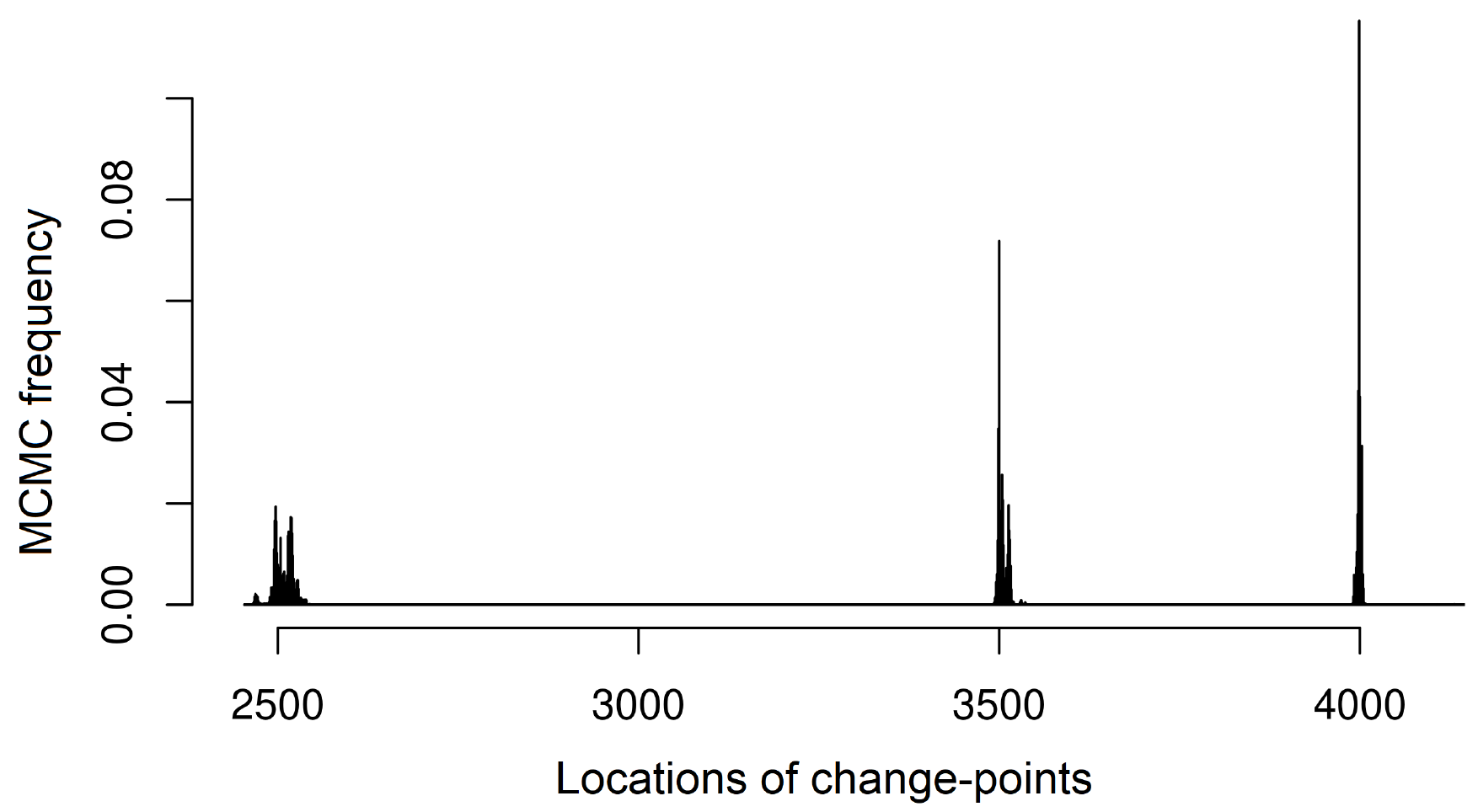}
    \caption{Simulated data.
	Left:
	MCMC histogram of the posterior distribution
of the number of change-points. 
	Right:
	MCMC histogram of
	the posterior distribution of the change-point locations.
In both cases,  
$N=10^5$ MCMC iterations were performed, 
with the first 10,000 samples discarded as burn-in. }
\label{fig:sim}
\end{figure}

The MCMC sampler 
of Section~\ref{sec32} was run 
with $\ell_{\rm max}=5$ and $D=5$;
see Figure~\ref{fig:sim}.

The posterior distribution of the number of change-points 
identifies 
the correct value $\ell=3$ with overwhelming 
confidence.
Similarly, the posterior distribution of the change-point locations
consists of three narrow peaks centered at the true locations.
The MAP estimates for the change-point locations are 
$\widehat p_ 1 = 2497$, $\widehat p _ 2 = 3500$, 
$\widehat p _3 =3999$.
Zoomed-in versions, showing the posterior distribution 
of the location of
the three change-points, are shown in
Figure~\ref{fig:MAP_example_me}
in Appendix~\ref{app:additional}.

\subsubsection{Comparisons to previous work}
\label{section:comparison}

In this slightly more extensive section,
we carefully compare the performance
of the \BCT-based method with that of one 
of the state-of-the-art algorithms used 
in probabilistic DNA sequencing,
the Bayesian HMM approach of \cite{totterdell:17}.
We first examine two 
datasets simulated from an HMM, and then
two datasets generated by an piecewise-homogeneous
variable-memory chain.

\medskip

\noindent
\textbf{HMM data.}
First we examine
two simulated datasets generated by the following
HMM whose parameters are the same as those 
used in \cite{totterdell:17}.
The hidden state process $\{S_n\}$ is a first-order 
Markov chain with values in $\{1,2\}$ and transition matrix,
$$    
\Lambda = \begin{bmatrix}
0.995 & 0.005  \\
0.005 & 0.995  \\
\end{bmatrix}.
$$
The observed process $\{X_n\}$ takes values
in $A=\{0,1,2,3\}$ and evolves as follows:
The conditional distribution of $X_n$ given
the value of the current state $S_n=k$
and the preceding value of $X_{n-1}=i$,
is generated according
to the transition matrix
$P^{(k)}=(P_{ij}^{(k)})$, where, for $k=1,2$:
\begin{align*}
P^{(1)} = & \begin{bmatrix}
0.20 & 0.30 & 0.30 & 0.20  \\
0.22 & 0.38 & 0.07 & 0.33  \\
0.23 & 0.27 & 0.32 & 0.18   \\
0.19 & 0.31 & 0.29 & 0.21 \\
\end{bmatrix} &
P^{(2)} & = \begin{bmatrix}
0.35 & 0.15 & 0.15 & 0.35  \\
0.37 & 0.13 & 0.13 & 0.37  \\
0.32 & 0.18 & 0.10 & 0.40   \\
0.35 & 0.20 & 0.20 & 0.25 \\
\end{bmatrix}.
\end{align*}
Both datasets have length $n=1000$. 
The first has a single change-point (that is,
a single state transition) at $p_1^*=412$, 
while the second one has three, at
$p_1^*=83$,
$p_2^*=657$,
and $p_3^*=862$.

Following \cite{boys:04}, 
in the Bayesian HMM approach of 
\cite{totterdell:17},
independent Dirichlet priors are placed on each 
row of the transitions matrices, 
so that
the posterior distribution of the parameters
$P^{(1)}$, $P^{(2)}$ and $\Lambda$ can easily
be obtained. Then the posterior distribution
of the state sequence is estimated through
a forward-backward algorithm.

\begin{figure}[!ht]
\vspace*{-0.14in}
\centering
  \includegraphics[width=5.3in]{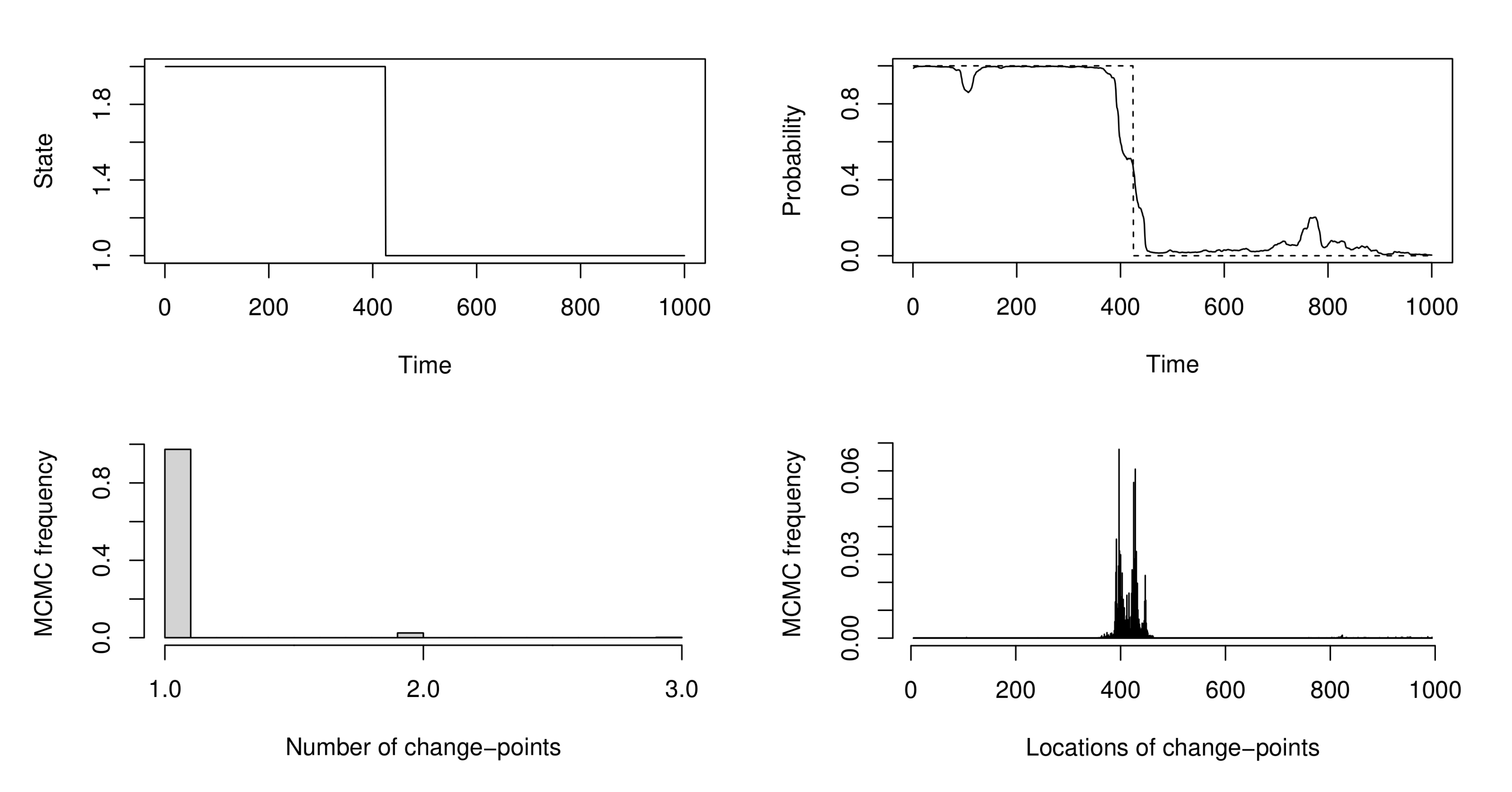}
\caption{Top left: Hidden state sequence with a change-point 
at $p_1^*=412$. Top right: Estimated 
segment probability, $\pi(S_k = 1| x)$, $1\leq k\leq n$;
true state sequence shown as the dashed line.
Bottom: MCMC histograms of the posterior distribution of the number 
of change-points (left) and of the posterior distribution of 
the change-points locations (right).}
\label{fig:totterdell_comp1}
\end{figure}

\begin{figure}[!ht]
\centering
  \includegraphics[width=5.2in]{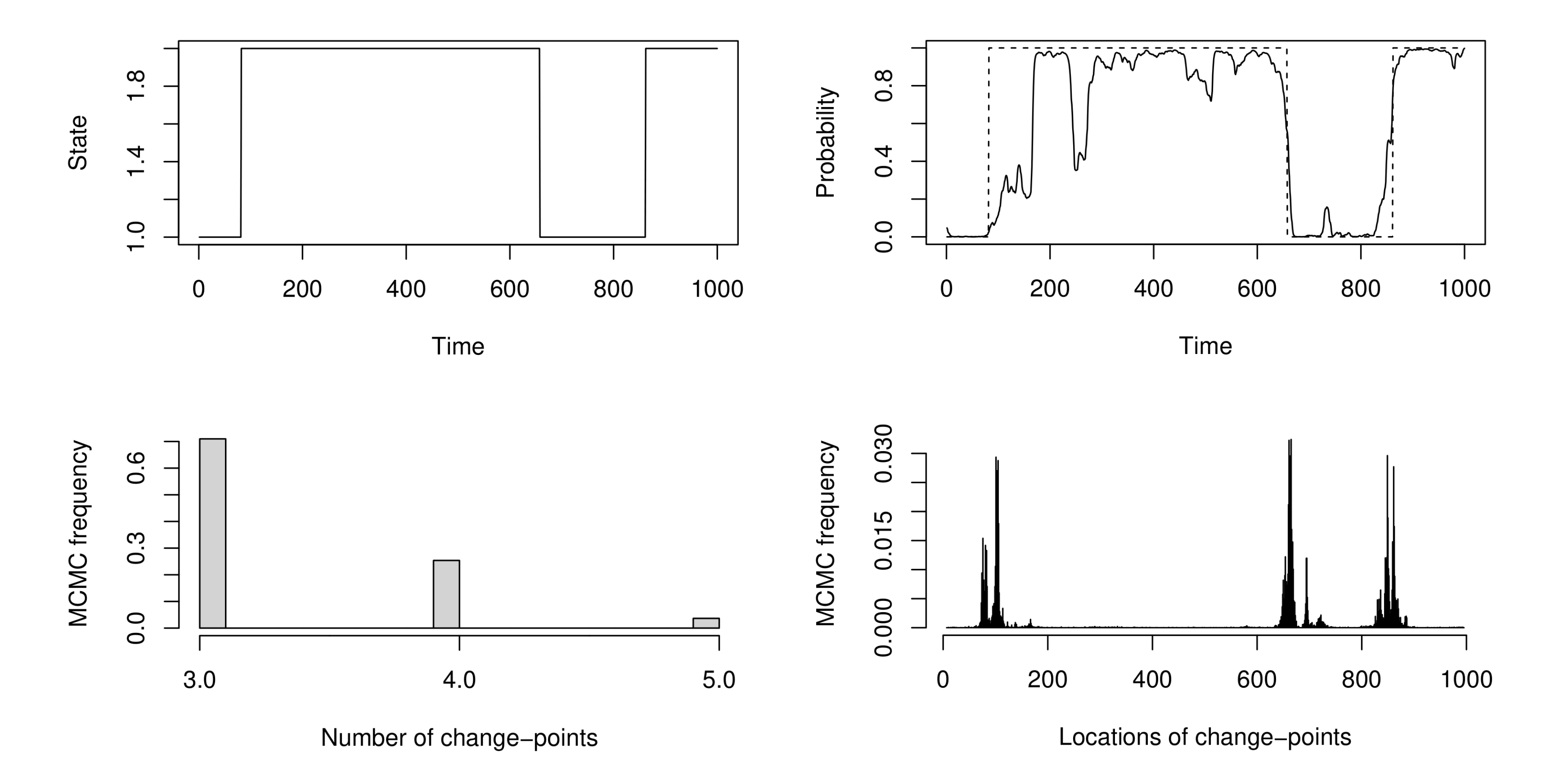}
\caption{Top left: Hidden state sequence with change-points 
at $p_1^*=83$, $p_2^*=657$, $p_3^*=862$. Top right: Estimated 
segment probability, $\pi(S_k = 1| x)$, $1\leq k\leq n$; true
state sequence shown as the dashed line.
Bottom: MCMC histograms of the posterior distribution of the number 
of change-points (left) and of the posterior distribution of 
the change-points locations (right).}
\label{fig:totterdell_comp3}
\end{figure}

For each simulated dataset, this Bayesian HMM algorithm was 
ran for $110,000$ iterations by initialising the 
parameters as described in \cite{totterdell:17}.
Our proposed change-point detection algorithm was ran for 
$110,000$ MCMC iterations,
with $\ell_{\rm max} = 5$ and $D = 3$. The results are shown
in Figures~\ref{fig:totterdell_comp1} and~\ref{fig:totterdell_comp3}.

Despite the fact that the data were generated by a model
that does not belong to the class of piecewise-homogeneous
variable-memory chains, the \BCT-based algorithm performs
essentially as well as the Bayesian HMM method
in identifying the number and locations of the
change-points. In addition, the \BCT-based algorithm provides
quantitative confidence estimates in the results obtained,
and it also more flexible in terms of prior knowledge: 
The Bayesian HMM approach requires the specification 
of the exact number of hidden states,
whereas the \BCT-based method only needs the specification 
of the much less critical parameter $\ell_{\rm max}$.

Finally, due to the efficiency of the 
\CTW\ and MCMC sampling algorithms, the proposed method
algorithm was
faster than the Bayesian HMM algorithm by
at least 2 orders of magnitude; 
see the first two rows of Table~\ref{table:running_time} in 
Appendix~\ref{section:running_time_tables}.

\medskip


\noindent
\textbf{Variable-memory chain data.} 
Next, we examine two data sets
generated by piecewise-homogeneous
variable-memory chains. Each dataset
has length $n=2000$ samples,
taking values in 
$A = \{0,1,2\}$,
with a single change-point 
at $p_1^*=1000$.
In each case, the two
segments 
are generated by variable-memory chains
with respect to the 
two models shown in Figure~\ref{fig:A4};
the associated parameters are given 
in Appendix~\ref{app:params}. 
In Figure~\ref{fig:totterdell_comp_new}, the 
results of the 
Bayesian HMM method
are compared with those of the 
\BCT-based algorithm with $\ell_{\rm max} = 3$ and  
$D = 5$.

\begin{figure}[!ht]
\centering
\includegraphics[width=5.2in]{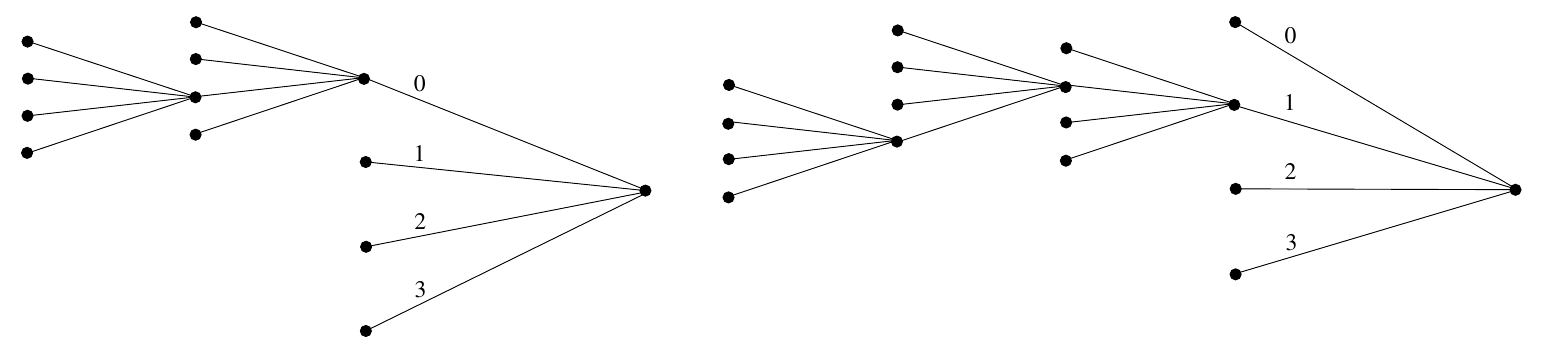}
\caption{The models used for the 
variable-memory Markov chain 
datasets in Section~\ref{section:comparison}.}
\label{fig:A4}
\end{figure}

\begin{figure}[!ht]
\centering
\includegraphics[width=5.0in]{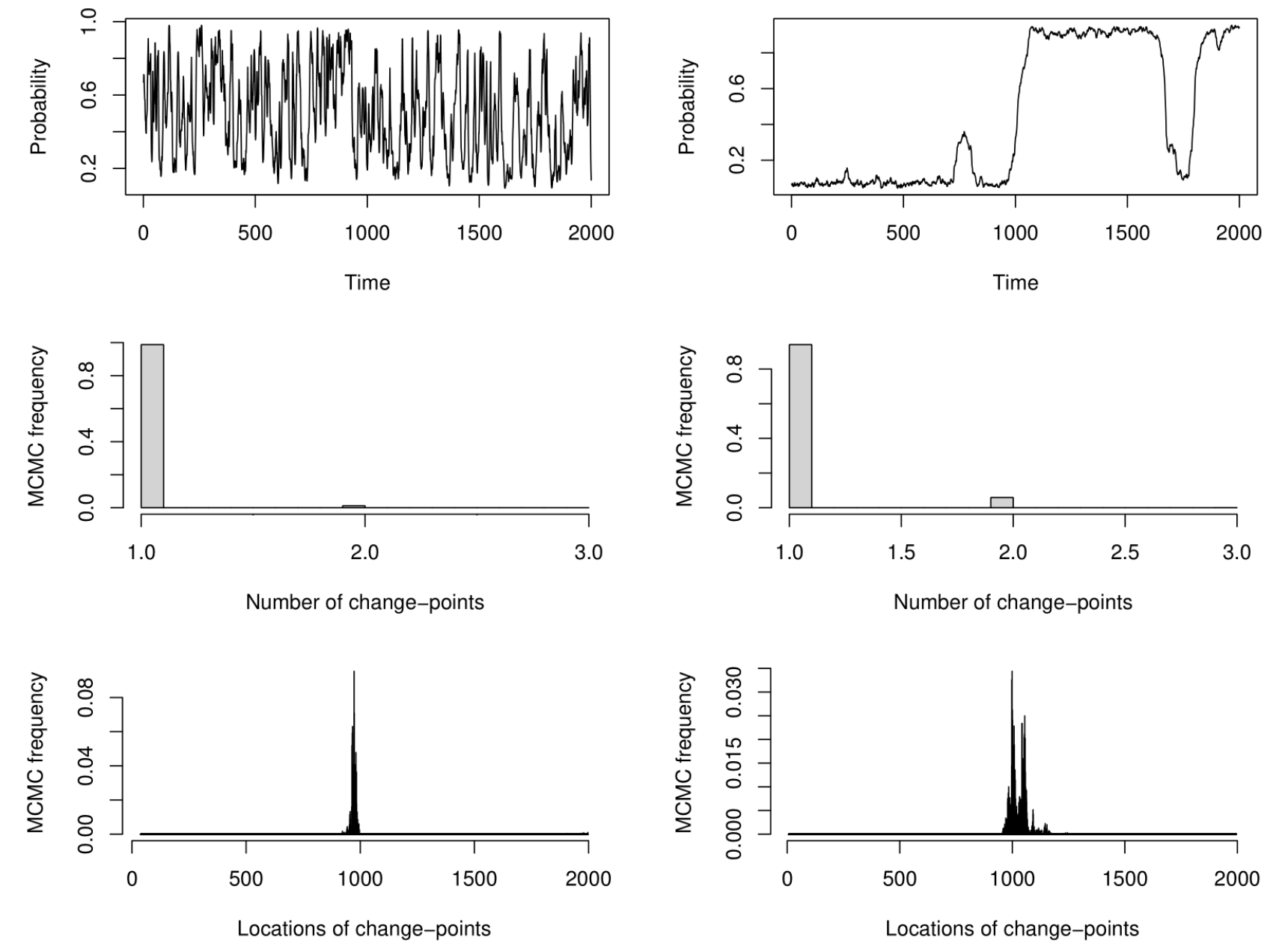}
\caption{Results of the Bayesian HMM approach 
and the \BCT-based change-point detection 
method, on two simulated datasets 
(one in each column)
generated from variable-memory Markov chains,
with a single change-point at $p_1^*=1000$.
For both algorithms, 
the same number $N = 110, 000$ of iterations were performed.}
\label{fig:totterdell_comp_new}
\end{figure}


For the first dataset (left column),
the empirical HMM-estimated probability 
of being in state~$1$ gives no information on 
the location of the change-point, whereas 
the \BCT-based algorithm both correctly identifies 
the presence of a single change-point
and provides an accurate estimate
of its location 
in the vicinity of positions 950-1000.
For the second dataset (right column), 
the empirical HMM-estimated probability 
is smoother and correctly indicates the 
location of the change-point, but it
also incorrectly identifies two non-existent 
changes. The \BCT-based method again
correctly identifies the 
presence of a single change point
as well as its location.

Unlike the robustness demonstrated by the
\BCT-based method on HMM data, 
the Bayesian HMM method appears to not be 
robust when used on data generated by a process
outside the HMM model class.
Also, as before, the 
running time of the Bayesian HMM algorithm
was found to be at least 100 times
longer than that of 
the \BCT-based algorithm; see 
Table~\ref{table:running_time} in
Appendix~\ref{section:running_time_tables}.

\subsubsection{Bacteriophage lambda}
\label{s:bl}

Here we revisit the $48,502$ base-pair-long genome of the 
bacteriophage lambda virus \citep{sanger:82}, available 
as sequence NC\_001416.1 at the GenBank database.
This dataset is often used as a benchmark for comparing 
different segmentation algorithms 
\citep{churchill:89,churchill:92,braun:98,braun:00,li:01,boys:04,gwadera:08}. 
Due to the high complexity of this genome,
which consists of 73 different genes, 
previous approaches often give different results 
on both the number and locations of change-points,
although all of them have been found to be biologically reasonable.
For example, \cite{gwadera:08} identify a total of~4 change-points, 
\cite{li:01,boys:04} identify~5, 
\cite{churchill:92,braun:98} identify~6, 
and \cite{braun:00} identify~8 change-points. 
These differences make it difficult to 
compare the performance of different methods 
quantitatively, but judging the biological relevance of the 
identified segments is still crucial. In order to do so, 
we will refer to the findings in the
standard biological reference work of \cite{liu:13}.

\begin{figure}[!ht]
\hspace{0.3in}
    \includegraphics[height=4.7cm]{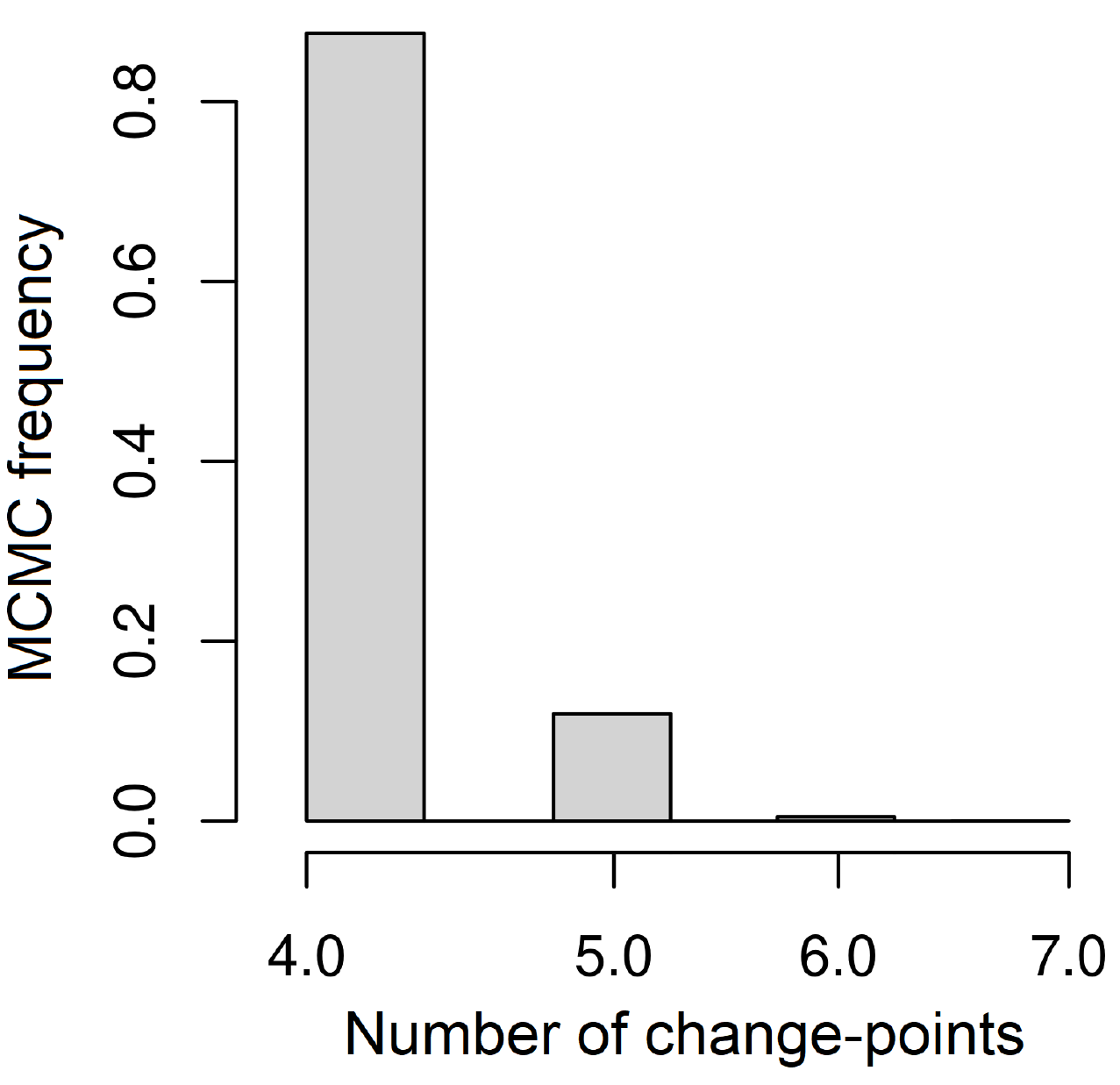}

\vspace*{-1.95in}

\hspace{2.2in}
    \includegraphics[height=4.8cm]{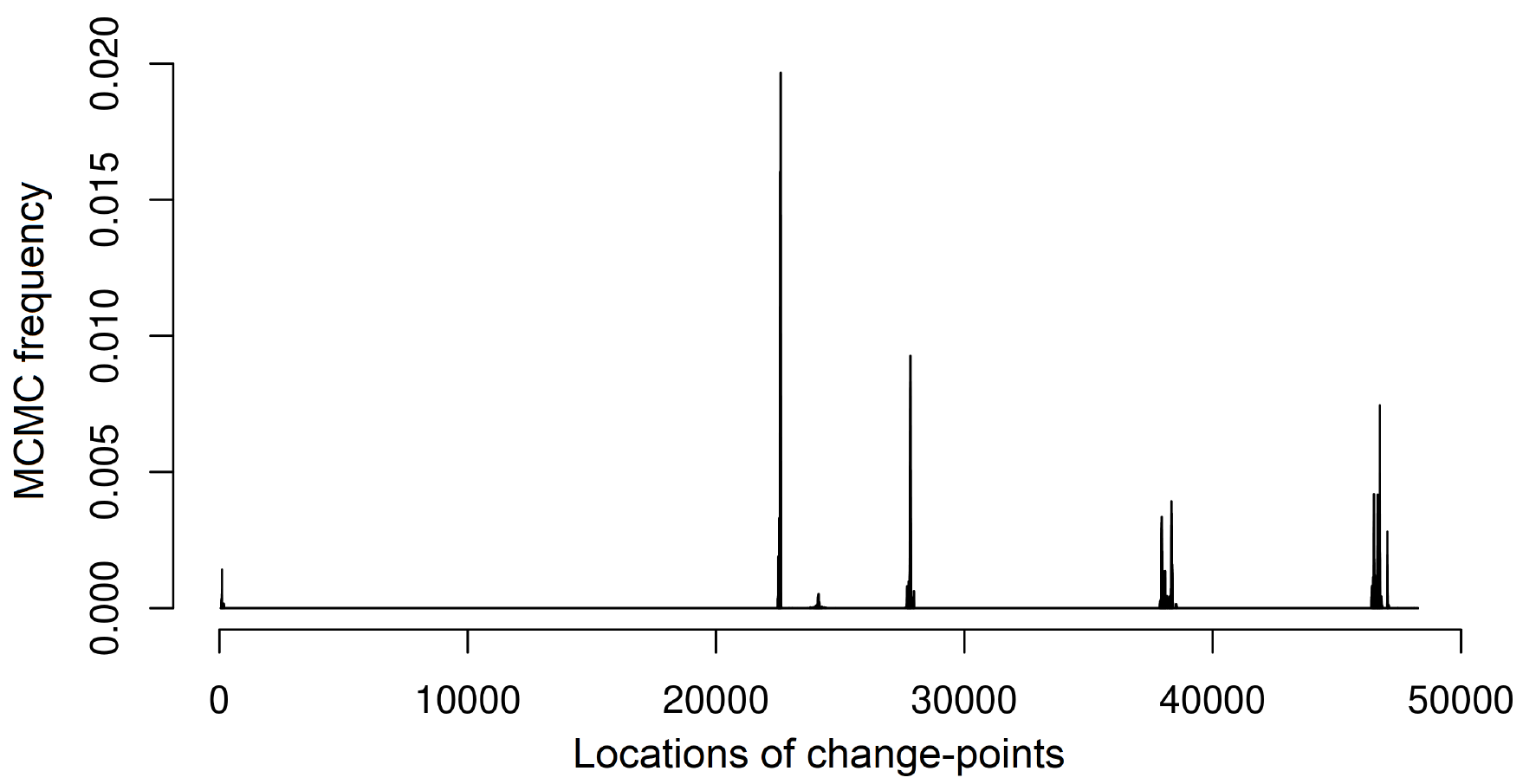}
\caption{Posterior distribution of the number and 
location of change-points
in the bacteriophage genome.
Left: MCMC histogram
of the posterior distribution of the number of change-points.
Right: MCMC
histogram of the posterior distribution of the change-point locations.
In both cases, $N=700,000$ MCMC samples were obtained,
with the first 70,000 discarded as burn-in.}
\label{fig:phage}
\end{figure}

The MCMC sampler 
of Section~\ref{sec32}
was run 
with $D=10$ and $\ell_{\rm max}=10$.
The resulting MCMC histogram approximation
of the posterior distribution over the number of 
change-points (Figure~\ref{fig:phage}, left plot) suggests 
that,
with very high probability, there are either $\ell=4$ or~$5$ 
change-points, with $\ell=4$ being over seven times 
more likely than $\ell=5$.

The posterior distribution of the change-point locations 
(Figure~\ref{fig:phage}, right plot) clearly identifies 
four significant locations, as well as two more,
around positions 100 and 24,000,
with significantly smaller weights.
Zoomed-in versions, showing the posterior 
distribution around
each of the four main locations are given
in Figure~\ref{fig:phage_4_map} in 
Appendix~\ref{app:additional}. 
The resulting MAP estimates for the change-point locations
are shown in Table~\ref{table:tables_phage}, where they are 
also compared with the biologically
``true'' change-points according to 
\cite{liu:13},
and with the estimates 
provided by one of the most reliable,
state-of-the-art
methods that have been applied to this data
\citep{braun:00}. It is noted that the high computational 
requirements associated with this example make the use of 
the software by \cite{totterdell:17} prohibitive in this case: 
Compared to the datasets 
considered in Section~\ref{section:comparison}, the present dataset is two orders of magnitude larger, and one would need to consider at least $r = 5$  hidden states.  

According to \cite{liu:13}, the segments we identify 
correspond to a biologically meaningful and important partition 
in terms of gene expression. The first segment (1-22607) starts at 
the beginning of the genome and ends very close to the start of gene ``ea47" (position $22686$), which signifies the end of the ``late operon" and the beginning of the leftward ``early operon'', both of which play an important role in transcription. The second segment (22608-27832) essentially consists of the region ``b2", an important region containing the three well-recognized ``early" genes ``ea47", ``ea31" and ``ea59". The third segment (27833-38340) ends very close to the end of gene ``cro" (position $38315$), which is the start of the rightward ``early operon" that is also essential in transcription. Lastly, the fourth segment (38341-46731) ends very close to the end (position $46752$) of gene ``bor", one of the major
genes being translated.

\begin{table}[!ht]
\centering
\begin{tabular}{cccc}
\hline
 Gene  & True & \BCT & \cite{braun:00} \\ \hline
ea47 & 22686 & { \bf 22607} & 22544 \\
ea59 & 26973 & { \bf 27832 }& 27829 \\
cro & 38315 & { \bf 38340}  & 38029 \\
bor & 46752 & { \bf 46731} & 46528 \\
 \hline
\end{tabular}
\caption{Estimates of change-point locations in the bacteriophage lambda
genome.}
\label{table:tables_phage}
\end{table}

Compared with previous findings, our results are similar enough to be plausible, while the places where they differ are precisely in the identification of biologically meaningful features, potentially improving performance. 
Specifically, our estimates 
are
very close to those 
obtained by \cite{gwadera:08}, where an approach based on variable-memory 
Markov models is also used: Four change-points are
identified by \cite{gwadera:08},
and their estimated locations lie inside the credible regions 
of our corresponding posterior distributions.

Compared with the alternative Bayesian approach of \cite{boys:04}, the present 
method gives 
similar locations and also addresses some of the known limitations 
of the Bayesian HMM approach. First, it was noted 
earlier that the Bayesian HMM framework 
is sensitive to the assumed prior distribution 
on the hidden state transition parameters. In contrast, 
the \BCT-based framework avoid this problem 
by relying on the prior predictive likelihood (which 
marginalizes over all models and parameters), 
using a simple default value for the only
prior hyperparameter, $\beta$.
Also, as argued by \cite{gupta:04}, the assumption 
that all segments share the same memory length may
be problematic for the HMM approach. 
The \BCT-based methodology
requires no such assumptions and,
indeed, allows for the memory length
of each segment to vary arbitrarily.
For example, the MAP tree models obtained
by the \BCT~algorithm in each segment 
of this dataset have maximum depths 
$d=5, 1, 2, 3$ and $0$, respectively,
indicating that the memory length indeed
varies throughout the genome.

\subsubsection{El Ni\~{n}o}
\label{s:EN}

El Ni\~{n}o \citep{trenberth:97} is one of the most influential 
natural climate patterns on earth. It impacts ocean temperatures, 
the strength of ocean currents, and the local weather in South America. 
As a result, it has direct societal consequences on
areas including 
the economy \citep{cashin:17} and public health \citep{kovats:03}.
Moreover, studying the frequency change of El Ni\~{n}o events 
can shed light on anthropogenic warming 
\citep{timm:99,cai:14,wang:19}. The dataset considered here is 
a binary time series that consists of $495$ annual observations 
between 1525 to 2020 \citep{quinn:87}, with 0 representing the absence 
of an El Ni\~{n}o event and 1 indicating its presence; data for recent years 
are also available online through the US Climate Prediction Center, at: 
\url{https://origin.cpc.ncep.noaa.gov/products/analysis_monitoring/ensostuff/ONI_v5.php}.

\begin{figure}[!ht]
    \centerline{\includegraphics[width=14cm]{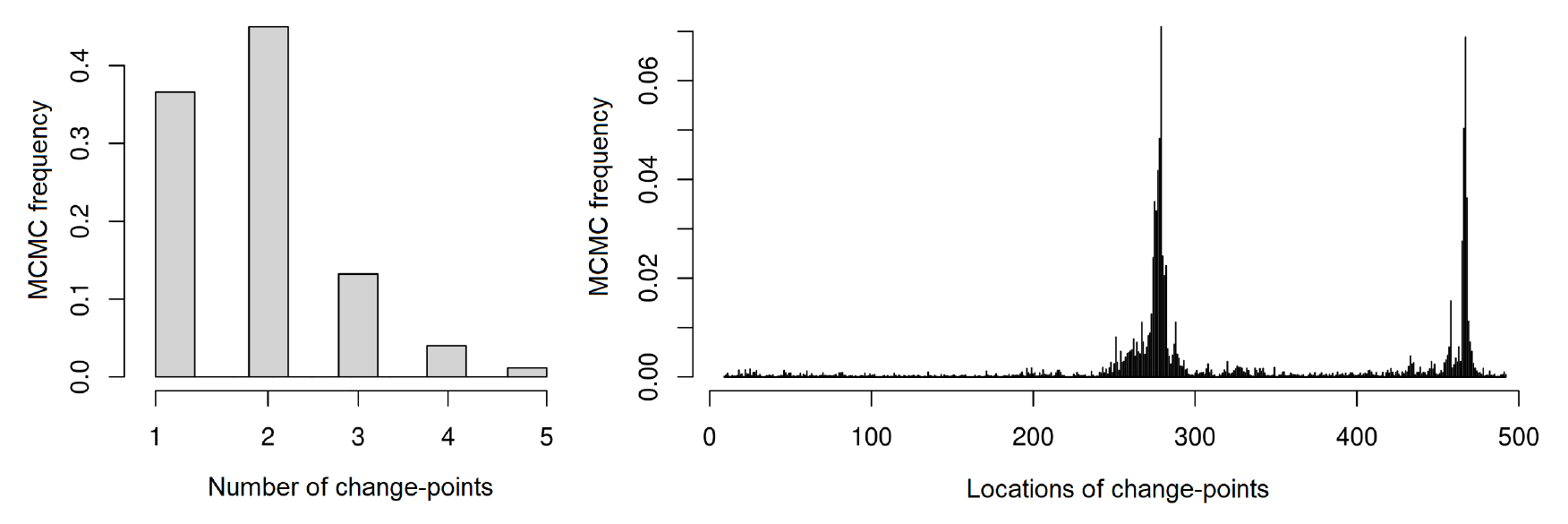}}
\caption{Posterior distribution of the number and 
location of the change-points 
for the El Ni\~{n}o dataset.
Left: MCMC histogram
of the posterior distribution of the number of change-points.
Right: MCMC
histogram of the posterior distribution of the change-point locations.
In both cases, $N=20,000$ iterations were performed and the first 2,000 
samples were discarded as burn-in.}
\label{fig:el_nino_figs}
\end{figure}

The MCMC sampler 
of Section~\ref{sec32}
was run with $D= 5$ and $\ell_{\rm max}=5$.
The resulting MCMC estimate 
of the posterior distribution on the number of change-points 
(left plot in Figure~\ref{fig:el_nino_figs}) suggests that the most likely 
value is $\ell=2$, with $\ell=1$ being a close second. The posterior 
distribution of
the change-point locations (right plot in Figure~\ref{fig:el_nino_figs}) 
also shows two clear peaks; zoomed-in versions
showing the location of each change-point separately 
are given in Figure~\ref{fig:nino_MAP}
in Appendix~\ref{app:additional}. 

The MAP estimates of the two locations 
are at $\widehat p_1 =278$ and $\widehat p _2 = 467$, 
corresponding to historically meaningful
events during the years 1802 and 1991, respectively:
The first change-point can be attributed to advancements in recording meteorological events, as prior to 1800 only the strong and extreme events were 
recorded \citep{quinn:87}. The second change-point in the early 1990s likely 
indicates 
a response to greenhouse warming, which is expected to increase both the frequency and the intensity of El Ni\~{n}o events \citep{cai:14,timm:99}. 
Indeed, examining 
the first-order marginal
$\pi = (\pi(0), \pi(1))$ 
of the 
stationary distribution 
associated 
with the MAP tree model in each segment, 
we find that the frequency~$\pi(1)$ of the recorded El Ni\~{n}o events 
increases between consecutive segments,
from $\pi(1) = 0.14$ before 1802, to $ \pi(1)  = 0.35$
between 1802 and 1991, and finally
to $ \pi(1) = 0.54$ after 1991.

\section{Conclusions} \label{sec5}

In this work, a new hierarchical Bayesian framework 
was developed for change-point detection and segmentation
of discrete time series, based on the \BCT\ family
of models and algorithms.
First, a new model class 
of piecewise-homogeneous variable-memory Markov chains
was described, and then 
an associated collection of algorithmic tools 
for segmentation and change-point detection 
was introduced.

The new \BCT-based methods are practical MCMC algorithms 
that only sample from the posterior 
distribution of the parameters of interest: 
The number of change points and their locations. 
The fact that all other parameters and models can be 
integrated out (by employing the \CTW\ algorithm)
results in computationally efficient samplers that 
were found to be at least 100 times faster than other
state-of-the-art techniques we compared with.
These MCMC samplers incorporate uniform random jumps 
that identify the approximate positions of change-points, 
as well as short-range random-walk moves that explore the 
high-probability regions around them. 
This way, the entire state space is explored effectively, 
providing an accurate estimate of the desired 
posterior distribution. 

An important advantage of the proposed tools is that 
they are based on a principled Bayesian approach, 
leading to general-purpose and application-agnostic methods
that do not require any prior information on the nature 
of the data. They are robust with respect to the underlying
parameter choices and, unlike earlier Bayesian-HMM approaches,
they do not require additional prior information on the number
of underlying states.

In numerous experimental settings with simulated
and real-world data, the \BCT-based methods were found
to give results as good or better than state-of-the-art
techniques. Moreover, they naturally also provide quantitative 
uncertainty estimates for all the results obtained. 
In particular, they were very effective in DNA segmentation 
problems, which form a key class of applications. 

The main limitation of the \BCT-based methods is the same
as that of the \BCT\ framework upon which they are built, 
namely, that they are only effective with data that take
on only a small number of possible values, typically 
no larger than 15 or 20. Therefore, an important direction 
for further work is to consider extensions that would work
effectively with large (or continuous) alphabets, particularly 
in the case of natural language processing. 
Another limitation is that our methods are all ``offline''
and can only detect change-points after the entire time
series has been observe. Therefore, another interesting 
direction for future work is the development of sequential 
methods, particularly in connection with timely applications 
related to online security.

\def\cprime{$'$}





\newpage

\appendix

\centerline{\LARGE\bf Appendix}


\section{MCMC acceptance probability}
\label{app:ratio}

The ratio $r((\ell, \pp), (\ell', \pp'))$ in the
acceptance probability of the MCMC sampler in
Section~\ref{sec32} is given by:
$$
\frac{P(x|\boldsymbol{p}', \ell')}{P(x|\boldsymbol{p},\ell)} \times \frac{ {\displaystyle\prod_{j=0}^{\ell'}}(p_{j+1}' - p_j' -1)}{ {\displaystyle \prod_{j=0}^{\ell}}(p_{j+1} - p_j -1)} \times \left\{
    \begin{array}{ll}
        \frac{2(n-2)}{(n-3)(n-4)},\text{ if } \ell = 0;
          \\
          \\
       \frac{(n-3)(n-4)}{2(n-2)}, \text{ if } \ell=1, \ell'=0;
        \\
        \\
       \frac{3(2\ell_{\rm max}+1)(n-\ell_{\rm max}-1)}{(n-2\ell_{\rm max}-2)(n-2\ell_{\rm max}-1)}, \text{ if } \ell=\ell_{\rm max}-1\text{, } \ell'=\ell_{\rm max};
       \\
       \\
       \frac{(n-2\ell_{\rm max}-2)(n-2\ell_{\rm max}-1)}{3(2\ell_{\rm max}+1)(n-\ell_{\rm max}-1)}, \text{ if } \ell=\ell_{\rm max}\text{, } \ell'=\ell_{\rm max}-1;
       \\
       \\
       \frac{(n-2\ell-2)(n-2\ell-1)}{2(2\ell+1)(n-\ell-1)}, 
	\text{ if } \ell'=\ell-1,\;\ell\neq 1,\ell_{\rm max};
       \\
       \\
       \frac{2(2\ell'+1)(n-\ell'-1)}{(n-2\ell'-2)(n-2\ell'-1)},
	\text{ if } \ell'=\ell+1,\;\ell'\neq 1,\ell_{\rm max};
       \\
       \\
       1,\text{ otherwise}.
    \end{array}
\right.
$$

\section{Maximum number of change-points}
\label{s:ellmax}

Results of the experiments in the last part 
of Section~\ref{section:maximum_number}. In both cases,
$N=300,000$ MCMC samples were obtained, 
with the first $30\%$ of the samples being discarded as burn-in.

\begin{figure}[!ht]
\centering
\begin{subfigure}{0.45\textwidth}
  \centering
  \includegraphics[width=\linewidth]{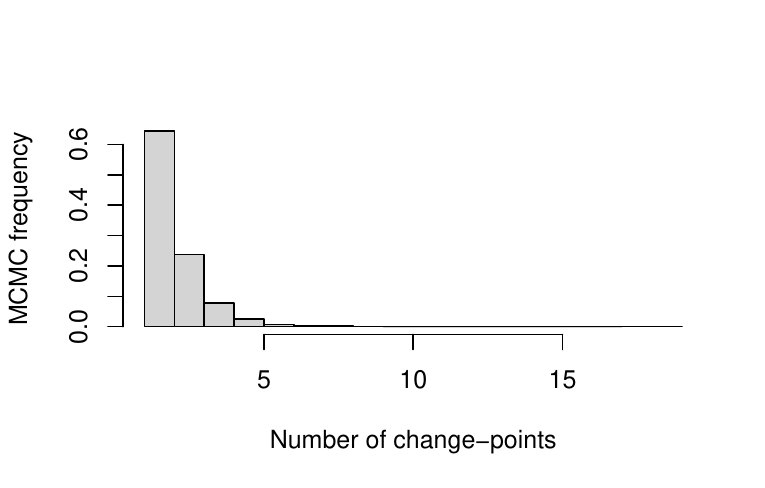}
\end{subfigure}
\begin{subfigure}{0.45\textwidth}
  \centering
  \includegraphics[width=\linewidth]{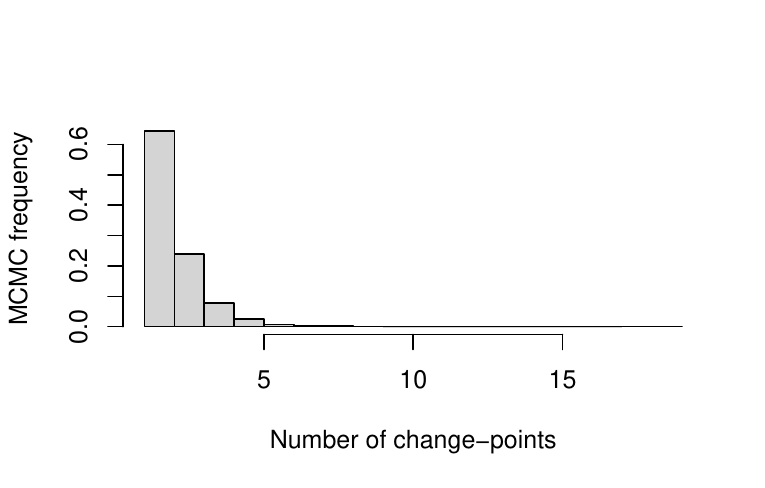}
\end{subfigure}
\begin{subfigure}{0.45\textwidth}
  \centering
  \includegraphics[width=\linewidth]{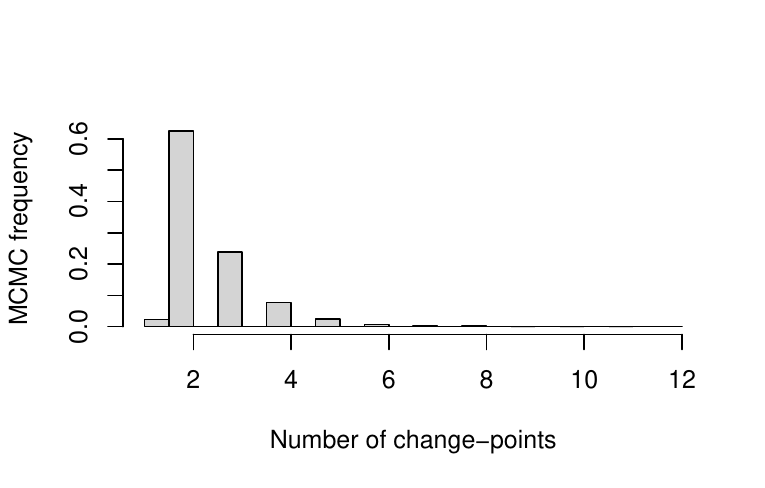}
\end{subfigure}
\begin{subfigure}{0.45\textwidth}
  \centering
  \includegraphics[width=\linewidth]{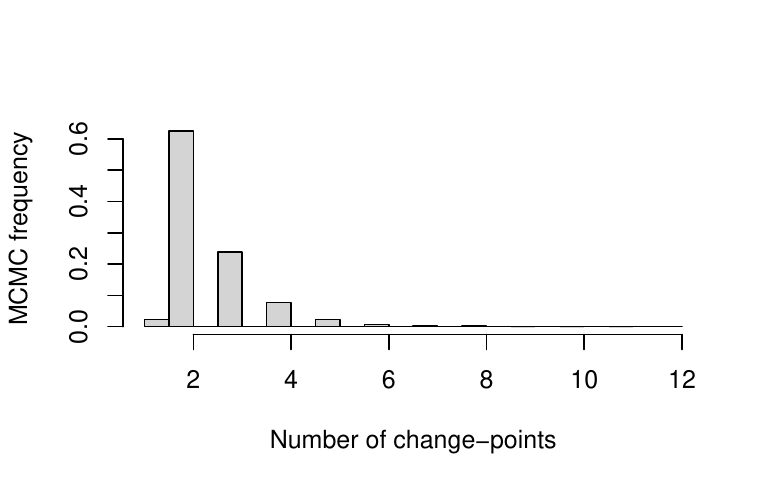}
\end{subfigure}
\caption{Posterior distribution 
of the number of change-points for 
dataset~1 in Section~\ref{section:maximum_number}.
From upper-left to lower-right, 
$\ell_{\rm max}= 25, 100, 250, 293$.}
\label{fig:lmax_data1}
\end{figure}

\begin{figure}[!ht]
\centering
\begin{subfigure}{0.45\textwidth}
  \centering
  \includegraphics[width=\linewidth]{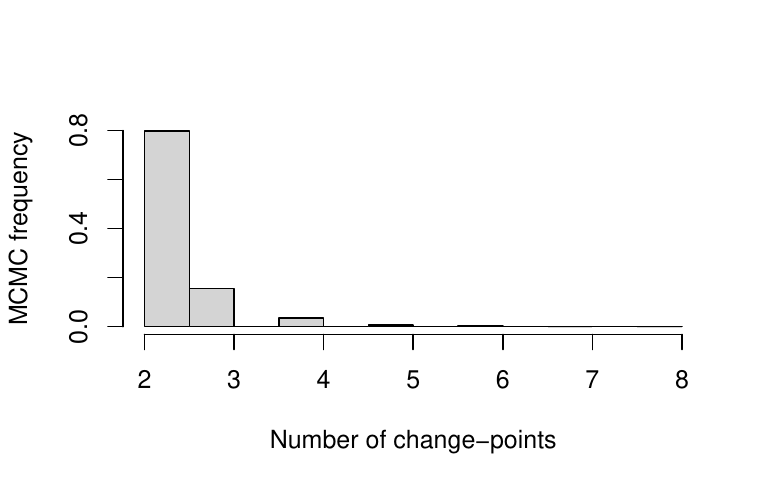}
\end{subfigure}
\begin{subfigure}{0.45\textwidth}
  \centering
  \includegraphics[width=\linewidth]{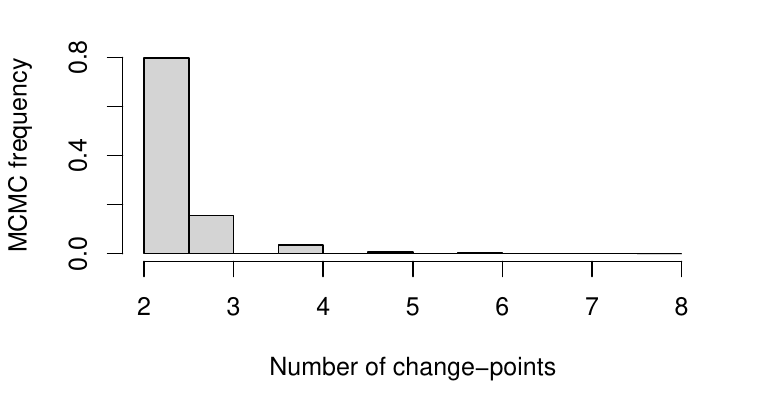}
\end{subfigure}
\begin{subfigure}{0.45\textwidth}
  \centering
  \includegraphics[width=\linewidth]{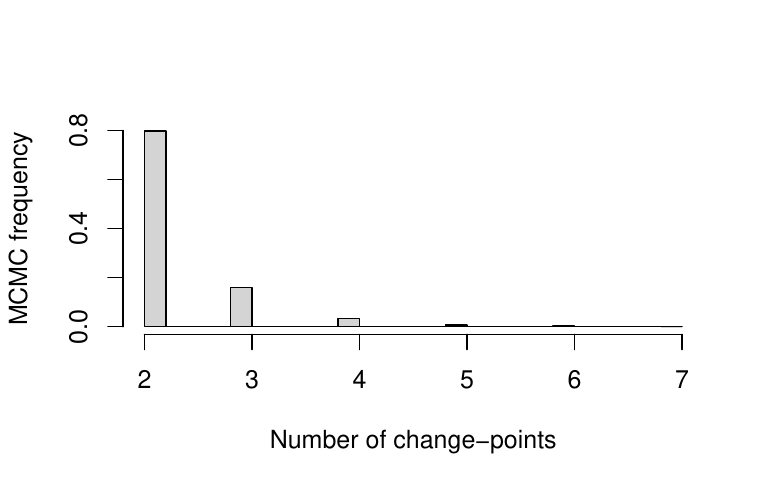}
\end{subfigure}
\begin{subfigure}{0.45\textwidth}
  \centering
  \includegraphics[width=\linewidth]{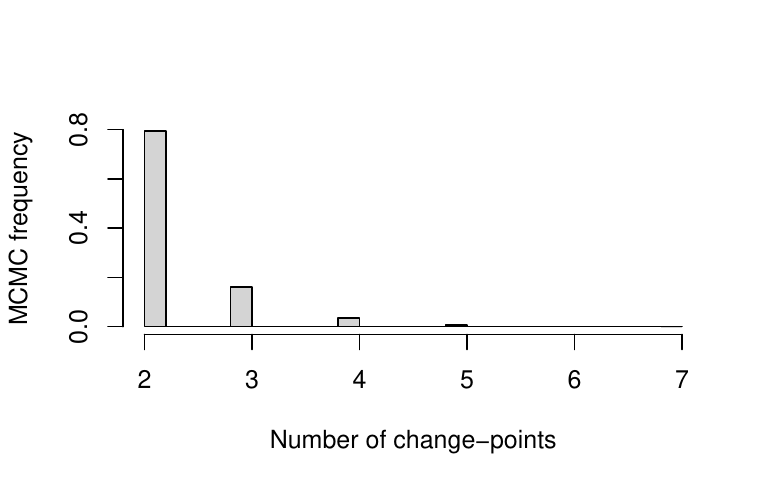}
\end{subfigure}
\vspace*{-0.2 cm}
\caption{Posterior distribution of the number of change-points for 
dataset~2 in Section~\ref{section:maximum_number}. 
From upper-left to lower-right, 
$\ell_{\rm max}=10, 150, 300, 600$.}
\label{fig:lmax_data2}
\end{figure}

\section{Empirical running times}
\label{section:running_time_tables}

\begin{table}[ht!]
\centering
\begin{tabular}{c|cccccccc}
\multicolumn{1}{c|}{} & \multicolumn{8}{c}{Dataset with $n=300$}     \\ \hline
$\ell_{\rm max}$ & 10   & 25 & 50  & 100 & 150 & 200 & 280 & 293 \\ \hline
Sampling time (seconds)    & 67.5 & 70 & 107 & 123 & 156 & 165 & 649 & 571\\
\hline
\multicolumn{1}{c|}{} & \multicolumn{8}{c}{Dataset with $n=900$} \\ \hline
$\ell_{\rm max}$ && 10  & 25 & 100 & 300 & 500 & 600 &  \\ \hline
Sampling time (seconds) && 114 & 143 & 138 & 759 & 3000 & 1791  &
\end{tabular}
\caption{MCMC sampling time for $N = 300,000$ iterations 
of the \BCT-based algorithm, for datasets~1 and~2
in Section~\ref{section:maximum_number}.}
\label{table:running_time_lmax1}
\end{table}

\begin{table}[ht!]
\centering
\begin{tabular}{c|cc}
\multicolumn{1}{c|}{} & \multicolumn{2}{c}{Running time (seconds)} \\ \hline
 & \BCT & Bayesian HMM \\ \hline
dataset 1 (HMM generated) & 58 & 7967 \\
dataset 2 (HMM generated) & 34 & 7108 \\
dataset 3 (VLMC generated) &  96  & 18510     \\
dataset 4 (VLMC generated) &  103  & 18030     \\
\end{tabular}
\caption{Running time for 110,000 iterations of 
the \BCT-based~algorithm
and the Bayesian HMM algorithm, for the four datasets 
in Section~\ref{section:comparison}.}
\label{table:running_time}
\end{table}

\section{Parameter values} 
\label{app:params}

\noindent
The parameters of the four models depicted
in Figure~\ref{fig:tree_models}
are given below.

\begin{table}[ht!]
\begin{tabular}{c|ccc}
\multicolumn{1}{c|}{Model 1}        & \multicolumn{3}{c}{Probability $P(j|s)$} \\ \hline
\multicolumn{1}{c|}{Context $s$} & $j=0$         & 1        & 2        \\ \hline
0                            & 0.3       & 0.4      & 0.3      \\
2                            & 0.5       & 0.3      & 0.2      \\
10                           & 0.2       & 0.5      & 0.3      \\
11                           & 0.1       & 0.4      & 0.5      \\
121                          & 0.7       & 0.2      & 0.1      \\
122                          & 0.4       & 0.2      & 0.4      \\
1200                         & 0.6       & 0.1      & 0.3      \\
1201                         & 0.3       & 0.5      & 0.2      \\
1202                         & 0.4       & 0.1      & 0.5     
\end{tabular}

\vspace{-2.09in}

\hspace{2.6in}
\begin{tabular}{c|ccc} 
\multicolumn{1}{c|}{ Model 2}  & \multicolumn{3}{c}{Probability $P(j|s)$} \\ \hline
\multicolumn{1}{c|}{Context $s$} & $j=0$         & 1        & 2        \\ \hline
0                            & 0.4       & 0.5      & 0.1      \\
2                            & 0.4       & 0.4      & 0.2      \\
10                           & 0.4       & 0.2      & 0.4      \\
11                           & 0.2       & 0.4      & 0.4      \\
12                           & 0.6       & 0.1      & 0.3     
\end{tabular}

\vspace{1.0in}

\begin{tabular}{c|ccc}
\multicolumn{1}{c|}{Model 3}        & \multicolumn{3}{c}{Probability $P(j|s)$} \\ \hline
\multicolumn{1}{c|}{Context $s$} & $j=0$         & 1        & 2        \\ \hline
0                            & 0.5       & 0.3      & 0.2      \\
1                            & 0.3       & 0.6      & 0.1      \\
2                            & 0.3       & 0.2      & 0.5   \\
\end{tabular}

\vspace{-0.97in}

\hspace{2.6in}
\begin{tabular}{c|ccc}
\multicolumn{1}{c|}{Model 4}        & \multicolumn{3}{c}{Probability $P(j)$} \\ \hline
\multicolumn{1}{c|}{} & $j=0$         & 1        & 2        \\ \hline
                       & 0.4       & 0.2      & 0.4   
\end{tabular}
\end{table}

\vspace{0.5in}


\noindent
The parameters of the two models depicted
in Figure~\ref{fig:A4}
are given below.

\begin{table}[ht!]
\begin{tabular}{c|cccc}
\multicolumn{1}{c|}{Model 1}        & \multicolumn{4}{c}{Probability $P(j|s)$} \\ \hline
\multicolumn{1}{c|}{Context $s$} & $j=0$         & 1        & 2     &3      \\ \hline 
1                            & 0.3       & 0.4      & 0.2   &  0.1 \\
2                            & 0.25      & 0.25     & 0.25  & 0.25 \\
3                            & 0.25      & 0.25     & 0.25  & 0.25  \\
00                           & 0.3       & 0.4      & 0.2   & 0.1  \\
01                           & 0.2       & 0.2      & 0.2   & 0.4  \\
03                           & 0.4       & 0.4      & 0.1   & 0.1  \\
020                          & 0.3       & 0.4      & 0.1   & 0.2  \\
021                          & 0.2       & 0.4      & 0.2   & 0.2  \\
022                          & 0.6       & 0.1      & 0.2   & 0.1  \\
023                         & 0.8       & 0.1      & 0     & 0.1
\end{tabular}

\vspace*{-5.8cm}
\hspace*{8.5cm}
\begin{tabular}{c|cccc} 
\multicolumn{1}{c|}{Model 2}        & \multicolumn{4}{c}{Probability $P(j|s)$} \\ \hline
\multicolumn{1}{c|}{Context $s$} &  $j=0$         & 1        & 2        & 3\\ \hline
0                            & 0.3       & 0.4      & 0.2      & 0.1\\
2                            & 0.4       & 0.2      & 0.3      & 0.1  \\
3                            & 0.25      & 0.25     & 0.25     & 0.25  \\
10                           & 0.1       & 0.1      & 0.2      & 0.6  \\
12                           & 0.2       & 0.1      & 0.5      & 0.2  \\
13                           & 0.4       & 0.1      & 0.4      & 0.1 \\
110                          & 0.2       & 0.3      & 0.3      & 0.2 \\
111                          & 0.3       & 0.4      & 0.1      & 0.2 \\
112                          & 0.7       & 0.1      & 0.1      & 0.1 \\
1130                         & 0.5       & 0.5      & 0        & 0 \\
1131                         & 0.4       & 0.3      & 0.1      & 0.2 \\
1132                         & 0.2       & 0.2      & 0.4      & 0.2 \\
1133                         & 0.3       & 0.2      & 0.2      & 0.3
\end{tabular}
\end{table}

\newpage

\section{Additional results}
\label{app:additional}

\begin{figure}[!ht]
\centerline{\includegraphics[width=12cm]{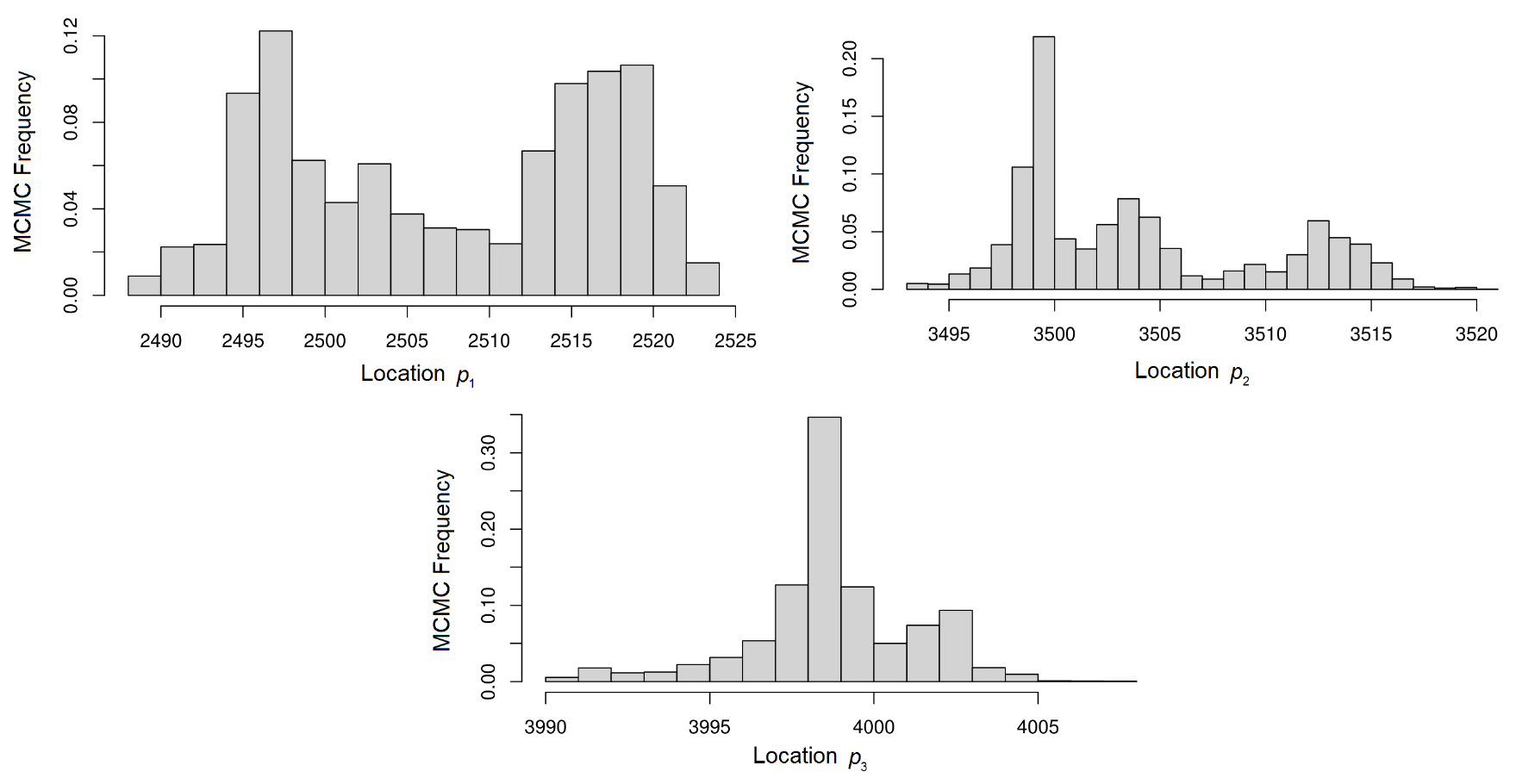}}
\caption{MCMC histograms
of the posterior distribution of each change-point location
in the simulated data example of Section~\ref{sec421}.}
\label{fig:MAP_example_me}
\end{figure}

\begin{figure}[!ht]
\centerline{\includegraphics[width=11cm]{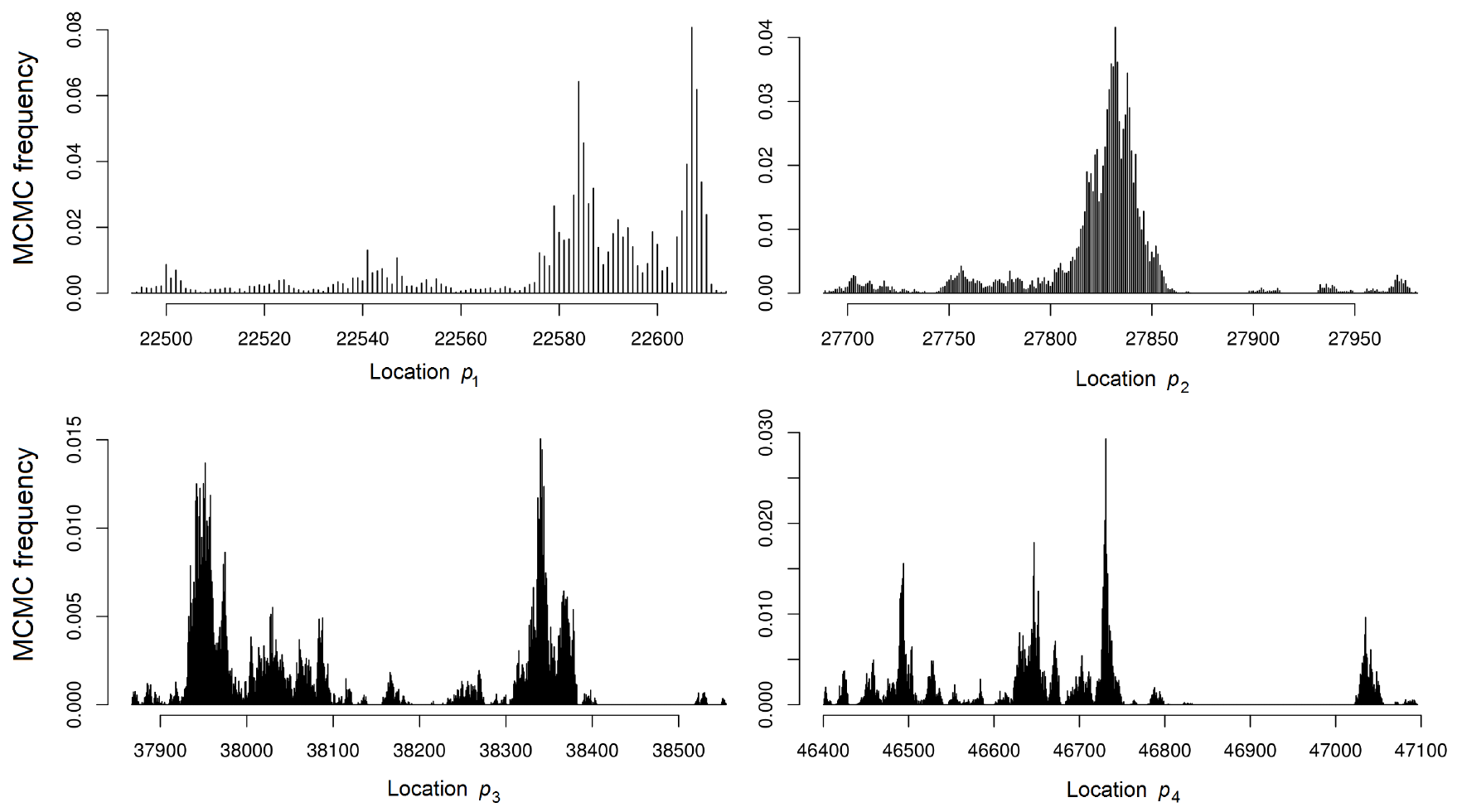}}
\caption{MCMC histograms
of the posterior distribution of each change-point location
in the bacteriophage lambda genome from Section~\ref{s:bl}.}
\label{fig:phage_4_map}
\end{figure}

\begin{figure}[!ht]
\centerline{\includegraphics[width=11cm]{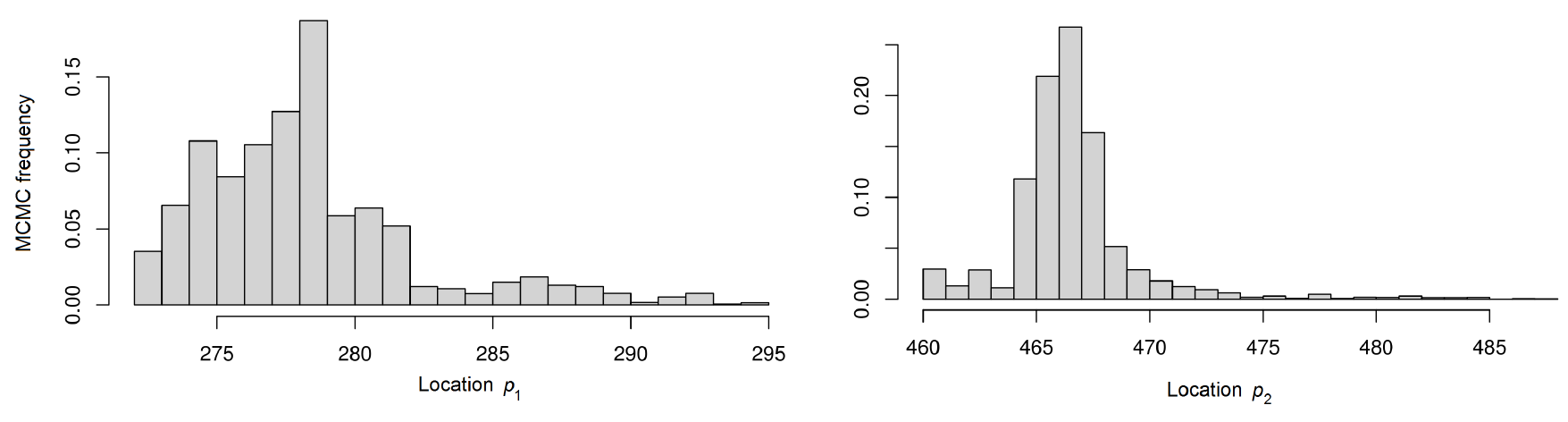}}
\caption{MCMC histograms
of the posterior distribution of each change-point location
in the meteorological data in Section~\ref{s:EN}.}
\label{fig:nino_MAP}
\end{figure}

\end{document}